\documentstyle[12pt]{article}
\newcommand{\be}{\begin{eqnarray}}
\newcommand{\ee}{\end{eqnarray}}

\begin{document}

\rightline{}

\vspace{1cm}

\begin{center}

\LARGE{Realistic model of the nucleon Spectral Function in few- and
many-nucleon systems}

\vspace{1.25cm}

\large{C. Ciofi degli Atti$^{(a)}$ and S. Simula$^{(b)}$}

\vspace{1cm}

\normalsize{$^{(a)}$Department of Physics, University of Perugia and\\
Istituto Nazionale di Fisica  Nucleare, Sezione di Perugia,\\ Via A. Pascoli,
I-06100 Perugia, Italy\\ $^{(b)}$Istituto Nazionale di Fisica Nucleare,
Sezione Sanit\'a,\\ Viale Regina Elena 299, I-00161 Roma, Italy}

\end{center}

\vspace{1cm}

\begin{abstract}

\indent By analysing the high momentum features of the nucleon momentum
distribution in light and  complex nuclei, it is argued that the
basic two-nucleon configurations  generating the structure of the nucleon
Spectral Function at high values of the nucleon momentum and removal energy,
can be properly described by a factorised ansatz for the nuclear wave
function, which leads to a nucleon Spectral Function in the form of a
convolution integral involving the momentum distributions describing the
relative and center-of-mass motion of a correlated nucleon-nucleon pair
embedded in the medium. The Spectral Functions of $^3He$ and infinite nuclear
matter resulting from the convolution formula and from many-body
calculations are compared, and a very good agreement in a wide range of
values of nucleon momentum and removal energy is found. Applications of the
model to the analysis of inclusive and exclusive processes are presented,
illustrating those features of the cross section which are  sensitive
to that part of the Spectral Function which is governed by short-range and
tensor nucleon-nucleon correlations.

\end{abstract}

\vspace{1cm}

\indent PACS number(s) : 21.10.Jx, 21.65.+f, 24.10.Cn, 27.20.+n

\newpage

\pagestyle{plain}

{\Large{\bf 1. Introduction}}

\vspace{0.5cm}

The nucleon Spectral Function $P(k,E)$ represents the joint probability to
find in a nucleus a nucleon with momentum $k \equiv |\vec{k}|$ and removal
energy $E$, and therefore it provides fundamental information on the dynamics
of the nucleon in the nuclear medium. Since the momentum and removal energy
dependences of $P(k,E)$ are governed by the single particle features of
nuclear structure as well as by the behaviour of the nuclear wave function at
short nucleon-nucleon ($NN)$ separations (see Ref. \cite{CSFS91}), the
relevance of the experimental and theoretical investigations of $P(k,E)$ is
clear.

\indent As far as the experimental investigation is concerned, it is well
known that within the impulse approximation the cross section for nucleon
knock-out processes is directly proportional to the nucleon Spectral Function,
and, although final state interactions and meson exchange currents can destroy
such a proportionality, several experiments have already provided relevant
information on the general features of $P(k,E)$, e.g.: i) the y-scaling
analysis \cite{CPS91} of inclusive $(e,e')$ experiments \cite{DAY} has
clarified the link between $P(k,E)$ and the nucleon momentum distribution
$n(k)$; ii) the high-resolution $A(e,e'p)X$ experiments have shown that the
shell-model occupation numbers can be substantially less than one, which is a
clear signature of the breakdown of the mean field picture \cite{EXP}; iii)
the exclusive $A(e,e'p)X$ experiments on $A =3$ and $A=4$ nuclei, performed
in kinematical regions corresponding to high excitation energies of the
residual nuclear system, have given the first direct signature of the effects
of $NN$ correlations on $P(k,E)$ \cite{MAR88,MAG94}; iv) the recent data on
the process $^{12}C(e,e'p)X$ have raised the question of multi-nucleon
emissions generated by $NN$ correlations \cite{MIT,NIK}. The above
experimental information clearly indicates that the single-particle strenght
is not concentrated uniquely at low values of $k$ and $E$ (as it is the case
within a mean field picture), but, due to short-range and tensor $NN$
correlations, is spread over a wide range of nucleon momenta and removal
energies.

\indent As far as the theoretical investigation of the nucleon Spectral
Function is concerned, the calculation of $P(k,E)$ for $A > 2$ requires the
knowledge of a complete set of wave functions for (A-1) interacting nucleons.
Thus, since the latter ones should be obtained from many-body calculations
using realistic models of the $NN$ interaction, the evaluation of $P(k,E)$
represents a formidable task. In case of $^3He$ the nucleon Spectral Function
has been obtained using three-body Faddeev \cite{FAD} or variational
\cite{CPS80} wave functions, whereas for $A = \infty$ the evaluation of
$P(k,E)$ has been performed using the orthogonal correlated basis approach
\cite{BFF} and perturbation expansions of the one-nucleon propagator
\cite{DIC}-\cite{FO92}. It should be mentioned that $P(k,E)$ has also been
obtained for $A=4$ \cite{MS91} using a plane-wave approximation for the final
states of the three-nucleon continuum. Finally, it is only recently that the
results of the investigation within a $G$-matrix perturbation theory of the
effects of the short-range and tensor $NN$ correlations on the $p$-wave
single-particle Spectral Function of $^{16}O$, have been reported \cite{MD94}.
Thus, microscopic calculations of the full nucleon Spectral Function of
light and complex nuclei are still called for, and it is for this reason that
the development of models of $P(k,E)$ is useful and necessary.

\indent In Ref. \cite{CSFS91} a model of the nucleon Spectral Function has
been proposed according to which, at high values of the nucleon momentum and
removal energy, $P(k,E)$  is expressed as a convolution integral of the
momentum distributions describing the relative and center-of-mass ($CM$)
motion of a correlated $NN$ pair. The basic assumption of the model is that
the high momentum and high removal energy parts of the nucleon Spectral
Function are generated by ground-state configurations in which two nucleons
are very close and form a correlated pair, whose $CM$ is, at the same time,
far apart from the other ($A-2$) nucleons. This means that the two nucleons
in the Bpair have large relative momenta (i.e., $k_{rel} \equiv |\vec{k}_1 -
\vec{k}_2 | / 2 > 0.3~GeV/c \sim k_F$), whereas the $CM$ momentum of the pair
is a low one ($k_{CM} \equiv |\vec{k}_1 + \vec{k}_2 | < 0.3~GeV/c$). It has
been shown \cite{CSFS91} that such a model satisfactorily reproduces the
Spectral Function of $^3He$ and nuclear matter calculated within many-body
approaches using realistic models of the $NN$ interaction. However, a
comprehensive derivation of the model for complex nuclei was not presented
in Ref. \cite{CSFS91}; thus, the aim of this paper is to provide the general
formulation of the convolution model and to apply it both to few-nucleon
systems and complex nuclei.

\indent The paper is organized as follows. In Section $2$ the definition of
$P(k,E)$ as well as its general form for  uncorrelated and correlated
many-nucleon systems are recalled; since our model for the Spectral Function
relies on some peculiar features of the nucleon momentum distributions and
their relationships with the Spectral Function, in Section 2 this matter will
be also discussed in some details. In Section $3$ the convolution model of
$P(k,E)$ sketched in Ref. \cite{CSFS91} is extended to any value of the mass
number $A$. In Section 4 the predictions of our model both for few-nucleon
systems and complex nuclei are presented and compared with available many-body
calculations. In Sections 5 and 6 some applications of our model to the
analysis of inclusive and exclusive quasi-elastic electron scattering are
presented. Finally, the summary and the conclusions are presented in Section
7.

\vspace{1cm}

{\Large{\bf 2. General theoretical framework: the Spectral Function and the
nucleon momentum distribution.}}

\vspace{0.5cm}

{\large{\bf 2.1 General definitions}}

\vspace{0.5cm}

\indent The nucleon Spectral Function $P(k,E)$ gives the joint probability to
find in a nucleus a nucleon with momentum $k$ and removal energy $E$. Since
the latter is defined as $E \equiv |E_A| - |E_{A-1}| + E_{A-1}^*$ (
$E_{A-1}^*$ being the (positive) excitation energy of the system with ($A-1$)
nucleons measured with respect to its ground-state, and $E_A$ ($E_{A-1}$)
the binding energy of the nucleus $A$ ($A-1$)), the Spectral Function also
represents the probability that, after a nucleon with momentum $\vec{k}$ is
removed from the target, the residual ($A-1$)-nucleon system is left with
excitation energy $E_{A-1}^*$. Adopting a non-relativistic Schroedinger
description of nuclei, the nucleon Spectral Function is defined as follows
 \be
    P(k,E) = {1 \over 2 J_0 + 1} \sum_{M_0 \sigma} ~ \langle \Psi_A^0
    | a_{\vec{k}, \sigma}^{\dagger} ~  \delta[E - (H - E_A)] ~ a_{\vec{k},
    \sigma} | \Psi_A^0 \rangle
    \label{2.1.1}
 \ee
where $k \equiv |\vec{k}|$, $a_{\vec{k}, \sigma}^{\dagger}$ ($a_{\vec{k},
\sigma}$) is the creation (annihilation) operator of a nucleon with momentum
$\vec{k}$ and spin projection $\sigma$; $\Psi_A^0$ is the intrinsic
eigenfunction of the ground-state of the nuclear Hamiltonian $H$ with
eigenvalue $E_A$ and total angular momentum (and its projection) $J_0$
($M_0$). Note that, being $\Psi_A^0$ an intrinsic wave function, the motion of
the center of mass of the residual system cannot contribute to Eq.
(\ref{2.1.1}). Note, moreover, that both $\vec{k}$ and $E$ are intrinsic
quantities; the former one is the internal momentum coniugate to the relative
distance $\vec{z}$ between the nucleon and the center of mass of the ($A-1$)
system, and the latter is  directly related to the intrinsic quantity
$E_{A-1}^*$ by  $E = E_{A-1} - E_A + E_{A-1}^* \equiv E_{min} +  E_{A-1}^*$,
where $E_{min} \equiv |E_A| -| E_{A-1}|$ is the (positive) minimum value of
the removal energy. Using the completeness relation $\sum_f | \Psi_{A-1}^f
\rangle \langle \Psi_{A-1}^f | = 1$ for the final states of the residual
system, one has
 \be
    P(k,E) \equiv {1 \over 2 J_0 + 1} \sum_{M_0 \sigma} ~ \sum_f \left |
    \langle \Psi_{A-1}^f | a_{\vec{k}, \sigma} \right | \Psi_A^0 \rangle |^2
    ~ \delta[E - (E_{A-1}^f - E_A)]
    \label{2.1.2}
 \ee
where $\Psi_{A-1}^f$ is the intrinsic eigenfunction of the state $f$ of the
Hamiltonian $H_{A-1}$ with eigenvalue $E_{A-1}^f \equiv E_{A-1} + E_{A-1}^*$.
Introducing the overlap integral $G_{f0}^{\sigma}(\vec{z})$ defined as
 \be
    G_{f0}^{\sigma}(\vec{z}) \equiv \int d\vec{x} ... d\vec{y} ~ \left [
    \chi_{\sigma}^{1/2} ~ \Psi_{A-1}^f(\vec{x} ... \vec{y}) \right ]^* ~
    \Psi_A^0(\vec{x} ... \vec{y}, \vec{z})
    \label{2.1.3}
 \ee
where $\chi_{\sigma}^{1/2}$ is the two-component Pauli spinor of the nucleon,
one has (cf. Ref. \cite{CPS80})
 \be
    P(k,E) = {1 \over (2 \pi)^3}{1 \over 2 J_0 + 1} \sum_{M_0 \sigma} ~
    \sum_f | \int d\vec{z} ~ e^{i\vec{k} \cdot \vec{z}} ~
    G_{f0}^{\sigma}(\vec{z})|^2 ~ \delta[E - (E_{A-1}^f - E_A)]
    \label{2.1.4}
 \ee

\vspace{1cm}

{\large{\bf 2.2 The nucleon Spectral Function for uncorrelated and correlated
many-nucleon systems}}

\vspace{0.5cm}

\indent From Eq. (\ref{2.1.4}) it follows that the nucleon Spectral Function
can be represented in the following form (cf. \cite{CLS90})
 \be
    P(k,E) = P_0(k,E) + P_1(k,E)
    \label{2.2.1}
 \ee
with
 \be
    P_0(k,E) \equiv {1 \over (2 \pi)^3} {1 \over 2 J_0 + 1} \sum_{M_0 \sigma}
    ~ \sum_{f<c} ~ \left | \int d\vec{z} ~ e^{i\vec{k} \cdot \vec{z}}
    ~ G_{f0}^{\sigma}(\vec{z}) \right |^2 ~ \delta[E - (E_{A-1}^f - E_A)]
    \label{2.2.2}
 \ee
 \be
    P_1(k,E) \equiv {1 \over (2 \pi)^3}{1 \over 2 J_0 + 1} \sum_{M_0 \sigma}
    ~ \sum_{f>c} ~ \left | \int d\vec{z} ~ e^{i\vec{k} \cdot \vec{z}}
    ~ G_{f0}^{\sigma}(\vec{z}) \right |^2 ~ \delta[E - (E_{A-1}^f - E_A)]
    \label{2.2.3}
 \ee
where $f<c$ ($f>c$) means that all final states of the residual system below
(above) the continuum threshold are considered, i.e. in $P_0(k,E)$ only final
states corresponding to the discrete spectrum of $H_{A-1}$ are included,
whereas all final states belonging to the continuum spectrum of $H_{A-1}$
contribute to $P_1(k,E)$.

\indent Let us now consider the predictions of the mean field approach for
the nucleon Spectral Function. If we consider the nucleus as an ensemble of
independent nucleons filling shell-model states $\alpha$ with momentum
distributions $n_{\alpha}$ and single-particle energies
$\varepsilon_{\alpha}$, one has \footnote[1]{In this paper the
normalization of the nucleon momentum distribution n(k) is chosen to be
$\int_0^{\infty} dk ~ k^2 ~ n(k) = 1$.}
 \be
    P_0^{(SM)}(k,E) = {1 \over 4 \pi A} ~ \sum_{\alpha} ~ A_{\alpha} ~
    n_{\alpha}^{(SM)}(k) ~ \delta[E - |\varepsilon_{\alpha}|]
    \label{2.2.4}
 \ee
and
 \be
    P_1^{(SM)}(k,E) = 0
    \label{2.2.5}
 \ee
In Eq. (\ref{2.2.4}) $A_{\alpha}$ is the number of nucleons in the state
$\alpha$ ($\sum_{\alpha} A_{\alpha} = A$) and the sum over $\alpha$ runs only
over hole states of the target, which means that the occupation probability
$S_{\alpha}$ is
 \be
    S_{\alpha}^{(SM)} \equiv \int_0^{\infty} dk k^2 n_{\alpha}^{(SM)}(k)
    = 1 ~ ~ ~ ~ & \mbox{for $\alpha < \alpha_F$} \nonumber \\
    = 0 ~ ~ ~ ~ & \mbox{for $\alpha > \alpha_F$}
    \label{2.2.6}
 \ee
In the limit $A \rightarrow \infty$, the nucleon Spectral Function for a
non-interacting Fermi gas is given by
 \be
    P_0^{(FG)}(k,E) & = & {3 \over 4 \pi k_F^3} ~ \theta(k_F - k) ~ \delta[E
    + {k^2 \over 2M}] \nonumber \\
    P_1^{(FG)}(k,E) & = & 0
   \label{2.2.7}
 \ee
which implies that the occupation probability for the single-particle
plane-wave states is equal to $1$ for $k \leq k_F$ and $0$ for $k> k_F$.

\indent The main effect of $NN$ correlations is to deplete states below the
Fermi level and to make the states above the Fermi level partially occupied.
By such a mechanism $P_0(k,E) \neq P_0^{SM}(k,E)$ and $P_1(k,E) \neq 0$.
Disregarding the finite width of the states below the Fermi level, the
(modified) shell-model contribution can be written as
 \be
    P_0(k,E) = {1 \over 4 \pi A} \sum_{\alpha < \alpha_F} ~ A_{\alpha} ~
    n_{\alpha}(k) ~ \delta[E - |\varepsilon_{\alpha}|]
    \label{2.2.8}
 \ee
where the occupation probability for hole states ($\alpha < \alpha_F$) is
 \be
    S_{\alpha}  = \int_0^{\infty} dk k^2 n_{\alpha}(k) < 1
    \label{2.2.9}
 \ee
In case of nuclear matter one has \cite{BFF}
 \be
    P_0(k,E) = {3 \over 4 \pi k_F^3} ~ Z(k) ~ \theta(k_F - k) ~ \delta[E
    + e_v(k)]
    \label{2.2.8.nm}
 \ee
where $Z(k)$ is the hole strenght and $e_v(k)$ is the hole single-particle
energy spectrum (in absence of $NN$ correlations $e_v(k) = k^2 / 2M$ and
$Z(k) = 1$, so that the Fermi gas Spectral Function (Eq. (\ref{2.2.7})) is
recovered). As for $P_1(k,E)$, its general expression reads as follows
 \be
    P_1(k,E) = {1 \over (2 \pi)^3} {1 \over 2 J_0 + 1} \sum_{M_0 \sigma} ~
    \sum_{f \neq \{\alpha < \alpha_F \}} \left | \int d \vec{z} ~ e^{i\vec{k}
    \cdot \vec{z}} ~ G_{f0}^{\sigma}(\vec{z}) \right |^2 ~ \delta[E -
    (E_{A-1}^f - E_{A-1})]
    \label{2.2.10}
 \ee
where more complex configurations of the ($A-1$)-nucleon system are included,
like, e.g., one-particle - two-hole states which can be reached when
two-particle - two-hole states in the target nucleus are considered. In such
a way, the main effects of $NN$ correlations on $P_1(k,E)$ is to generate
high momentum and high removal energy components.

\indent Before closing this subsection, let us consider explicitely the case
of nuclei with $A \leq 4$, for in this case the mean field picture underlying
Eq. (\ref{2.2.8}) is not a realistic one. In case of the deuteron there is
only one possible final state, because the residual (A-1) system is just a
nucleon; this means that $P_1(k,E)$ for $A=2$ is identically vanishing, even
in presence of $NN$ correlations. Thus, the dependence of $P(k,E)$ upon the
removal energy is simply given by a delta function. Explicitely, the nucleon
Spectral Function of the deuteron reads as follows
 \be
    P(k,E) = {1 \over 4 \pi} n^{(D)}(k) ~ \delta[E - E_{min}]
    \label{2.2.11}
 \ee
where $E_{min} = 2.226~MeV$ is the deuteron binding energy and $n^{(D)}(k)$
is the nucleon momentum distribution in the deuteron (see the next
subsection). In case of helium nuclei the residual ($A-1$)-nucleon system has
only one bound state, namely the ground state of the two- and three-nucleon
system for $A = 3$ and $A = 4$, respectively. Thus, Eq. (\ref{2.2.2}) yields
the nucleon Spectral Function associated to the ground-to-ground transition,
viz.
 \be
    P_0(k,E) & = & {1 \over (2 \pi)^3} {1 \over 2 J_0 + 1} \sum_{M_0 \sigma}
    ~ \left | \int d\vec{z} e^{i\vec{k} \cdot \vec{z}}
    ~ G_{00}^{\sigma}(\vec{z}) \right |^2 ~ \delta[E - E_{min}]
    \label{2.2.12}
 \ee
where $E_{min} \equiv |E_A| - |E_{A-1}|$ equals to $5.5~MeV$ and $19.8~MeV$
in case of $^3He$ and $^4He$, respectively. As far as $P_1(k,E)$ is concerned,
it is given by
 \be
    P_1(k,E) = {1 \over (2 \pi)^3} {1 \over 2 J_0 + 1} \sum_{M_0 \sigma}
    ~ \sum_{f \neq 0} \left | \int d \vec{r} ~ e^{i\vec{k} \cdot \vec{r}} ~
    G_{f0}^{\sigma}(\vec{r}) \right |^2 ~ \delta[E - (E_{A-1}^f - E_{A-1})]
    \label{2.2.13}
 \ee

\vspace{1cm}

{\large{\bf 2.3 The Spectral Function and the nucleon momentum distribution}}

\vspace{0.5cm}

The nucleon momentum distribution $n(k)$ is defined as follows
 \be
    n(k) \equiv {1 \over 2 \pi^2} \int d \vec{z} d \vec{z}' ~ e^{i \vec{k}
    \cdot (\vec{z} - \vec{z}')} ~ \rho(\vec{z}, \vec{z}')
    \label{2.3.1}
 \ee
where (omitting spin indices)
 \be
    \rho(\vec{z}, \vec{z}') \equiv \int d \vec{x} ... d \vec{y} ~ \left [
    \Psi_{A}^0(\vec{x} ... \vec{y}, \vec{z}) \right ]^* ~ \Psi_A^0(\vec{x}
    ... \vec{y}, \vec{z}')
    \label{2.3.2}
 \ee
is the off-diagonal one-body density matrix. The calculation of $n(k)$ is
simpler than the calculation of $P(k,E)$, for only the ground-state wave
function is required. The relation between $n(k)$ and $P(k,E)$ can be easily
obtained by inserting in Eq. (\ref{2.3.2}) the completenss relation $\sum_f |
\Psi_{A-1}^f \langle \rangle \Psi_{A-1}^f | = 1$ for the final states of the
residual system and by using Eq. (\ref{2.1.4}); one obtains the following
momentum sum rule
 \be
    n(k) = 4 \pi ~ \int_{E_{min}}^{\infty} dE ~ P(k,E)
    \label{2.3.3}
 \ee
Using the decomposition given by Eq. (\ref{2.2.1}), one has (cf. \cite{HE3},
\cite{CLS90})
 \be
    n(k) =  n_0(k) + n_1(k)
    \label{2.3.4}
 \ee
with
 \be
    n_0(k) \equiv 4 \pi ~ \int_{E_{min}}^{\infty} dE ~ P_0(k,E) =
    {1 \over 2 \pi^2} {1 \over 2 J_0 + 1} \sum_{M_0 \sigma} ~ \sum_{f =
    \{\alpha < \alpha_F \}} \left | \int d \vec{z} ~ e^{i\vec{k} \cdot
    \vec{z}} ~ G_{f0}^{\sigma}(\vec{z}) \right |^2
    \label{2.3.5}
 \ee
 \be
    n_1(k) \equiv 4 \pi ~ \int_{E_{min}}^{\infty} dE ~ P_1(k,E) =
    {1 \over 2 \pi^2} {1 \over 2 J_0 + 1} \sum_{M_0 \sigma} ~ \sum_{f \neq
    \{\alpha < \alpha_F \}} \left | \int d \vec{z} ~ e^{i\vec{k} \cdot
    \vec{z}} ~ G_{f0}^{\sigma}(\vec{z}) \right |^2
    \label{2.3.6}
 \ee
For $A = 3$ and $A = 4$, one gets (cf. Eq. (\ref{2.2.12}))
 \be
    n_0(k) = {1 \over 2 \pi^2} {1 \over 2 J_0 + 1} \sum_{M_0 \sigma} ~
    \left | \int d\vec{z} ~ e^{i\vec{k} \cdot \vec{z}} ~
    G_{00}^{\sigma}(\vec{z}) \right |^2
    \label{2.3.7}
 \ee
whereas for a complex nucleus, when $NN$ correlations are considered, one
has (cf. Eq. (\ref{2.2.8}))
 \be
    n_0(k) = {1 \over 4 \pi A} \sum_{\alpha < \alpha_F} ~ A_{\alpha} ~
    n_{\alpha}(k)
    \label{2.3.7bis}
 \ee
It can be seen therefore that in both cases there is a direct relationship
between $P_0(k,E)$ and $n_0(k)$, viz.
 \be
    P_0(k,E) = {1 \over 4 \pi} n_0(k) ~ \delta[E - E_{min}]
    \label{2.3.7ter}
 \ee
for a few-body system, and Eq. (\ref{2.2.8}) for a complex nucleus. On the
contrary, $P_1(k,E)$ cannot be expressed in terms of $n_1(k)$ both in case
of light and complex nuclei.

\indent The integration of $n_0(k)$ and $n_1(k)$ over the nucleon momentum
yields the spectroscopic factors $S_0$ and $S_1$, respectively, viz.
 \be
    S_0 & \equiv & \int_0^{\infty} dk ~ k^2 ~ n_0(k) = \sum_{\alpha <
    \alpha_F} ~ S_{\alpha} \nonumber \\
    S_1 & \equiv & \int_0^{\infty} dk ~ k^2 ~ n_1(k)
    \label{2.3.8}
 \ee
with $S_0 + S_1 = 1$. Within the mean field picture one has $S_0 = 1$ and $S_1
= 0$. When $NN$ correlations are considered, it turns out that the
spectroscopic factor of the ground-to-ground transition $S_0$ is equal to
$\simeq 0.65$ and $\simeq 0.8$ (implying $S_1 \simeq 0.35$ and $\simeq 0.2$)
for $^3He$ \cite{HE3,FAD} and $^4He$ \cite{HE4_1,MS91}, respectively, whereas
the average depletion of single-particle states below Fermi level in complex
nuclei is expected to be $S_0 \simeq 0.8$ in overall agreement with the
results obtained from high-resolution exclusive experiments \cite{EXP}. In
case of the infinite nuclear matter the calculation of Ref. \cite{NM} yields
$S_0 \simeq 0.75$, which means $S_1 \simeq 0.25$.

\indent The nucleon momentum distribution $n(k)$ has ben calculated for
several nuclei, ranging from the deuteron to nuclear matter \cite{HE3} -
\cite{ORD80}, using non-relativistic ground-state wave functions with
realistic models of the $NN$ interaction (like, e.g. the Reid Soft Core
interaction \cite{RSC}, the Paris potential \cite{PARIS} and the Argonne
$v14$ interaction \cite{ARG14}) and considering also three-nucleon forces (cf.
Ref. \cite{HE4_3}). A sample of the results of those calculations is reported
in Fig. 1, where it can be clearly seen that: i) the low-momentum part of
$n(k)$ ($k < 1.5 ~ fm^{-1}$) is almost totally exhausted by $n_0(k)$, which
means that $n(k)$ at low momenta is dominated by the single-particle features
of the nuclear structure; ii) the high-momentum tail of $n(k)$ ($k > 1.5 ~
fm^{-1}$) is entirely governed by $n_1(k)$, which at high momenta overwhelms
the contribution from $n_0(k)$ by several orders of magnitude; this means that
$n(k)$ at high momenta is governed by the short-range properties of nuclear
structure; iii) both at low and high momenta the results of the existing
calculations of $n(k)$ agree fairly well with the general trend of the
nucleon momentum distribution extracted from inclusive and exclusive
experiments.

\indent In Fig. 2(a) the nucleon momentum distributions calculated for complex
nuclei are directly compared with the one of the deuteron. It can be clearly
seen that the high-momentum tail of $n(k)$ at $k > 2 ~ fm^{-1}$ is similar
for all nuclei and it is essentially given by the nucleon momentum
distribution of the deuteron times an $A$-dependent scale factor. Therefore,
the ratio of $n^A(k)$ for a nucleus $A$ to $n^D(k)$ for the deuteron is
expected to exhibit a plateaux for $k > 2 ~ fm^{-1}$, as it indeed appears
in Fig. 2(b). The height of the plateaux turns out to be $\sim 2$ for the
proton in $^3He$, $\sim 4$ in $^4He$ and $^{16}O$, $\sim 4.5$ in $^{56}Fe$
and $\sim 5$ in nuclear matter. The similarity between the high momentum
parts of the momentum distributions for complex nuclei and the deuteron has
been firstly illustrated in Ref. \cite{ZAB78}.

\indent From the above considerations it follows that a simple model for the
nucleon momentum distribution $n(k)$ could read as follows
 \be
    n(k) \simeq n_0(k) = {1 \over 4 \pi A} \sum_{\alpha < \alpha_F} ~
    A_{\alpha} ~ n_{\alpha}(k) ~ ~ ~ ~ ~ ~ ~ ~ \mbox{for $k < \tilde{k}$}
    \nonumber \\
    n(k) \simeq n_1(k) = C^A ~ n_{deut}(k) ~ ~ ~ ~ ~ ~ ~ ~ \mbox{for $k >
    \tilde{k}$}
    \label{2.3.9}
 \ee
where $\tilde{k} \simeq 2.0 ~ fm^{-1}$. We have parametrized the results
of many-body calculations for $n_0(k)$ and $n_1(k)$, shown in Fig. 1,
using simple functional forms inspired by Eq. (\ref{2.3.9}) (cf. Appendix).

\indent In what follows we will make use of two important quantities, viz.
the relative and center-of-mass ($CM$) momentum distributions of a two-nucleon
cluster ($N_1N_2$), which are defined as follows
 \be
    n_{rel}^{N_1N_2}(\vec{k}_{rel}) = \int d \vec{k}_{CM} ~
    n^{N_1N_2} \left ( \vec{k}_{rel} + {\vec{k}_{CM} \over 2}, - \vec{k}_{rel}
    + {\vec{k}_{CM} \over 2} \right )
    \label{2.3.10}
 \ee
 \be
    n_{CM}^{N_1N_2}(\vec{k}_{CM}) = \int d \vec{k}_{rel} ~
    n^{N_1N_2} \left ( \vec{k}_{rel} + {\vec{k}_{CM} \over 2}, - \vec{k}_{rel}
    + {\vec{k}_{CM} \over 2} \right )
    \label{2.3.10bis}
 \ee
where $\vec{k}_{rel} \equiv$ $(\vec{k}_1 - \vec{k}_2) /2$ and $\vec{k}_{CM}
\equiv$ $\vec{k}_1 + \vec{k_2}$ are the relative and $CM$ momentum of the
pair $N_1 N_2$, respectively, with $\vec{k}_1$ and $\vec{k}_2$ being
measured from the $CM$ of the system. In Eq. (\ref{2.3.10})
$n^{N_1N_2}(\vec{k}_1, \vec{k}_2)$ is the two-nucleon momentum distribution
 \be
    n^{N_1N_2}(\vec{k}_1, \vec{k}_2) \equiv {1 \over (2 \pi)^6}  \int
    d\vec{z}_1 d\vec{z}_2 d\vec{z}'_1 d\vec{z}'_2 ~ e^{i \vec{k}_1 \cdot
    (\vec{z}_1 - \vec{z}'_1)} ~ e^{i \vec{k}_2 \cdot (\vec{z}_2 - \vec{z}'_2)}
    ~ \rho^{N_1N_2}(\vec{z}_1, \vec{z}_2; \vec{z}'_1, \vec{z}'_2)
    \label{2.3.11}
 \ee
where
 \be
    \rho^{N_1N_2}(\vec{z}_1, \vec{z}_2; \vec{z}'_1, \vec{z}'_2) \equiv \int
    d \vec{x} ... d \vec{y} ~ \left [ \Psi_{A}^0(\vec{x} ... \vec{y},
    \vec{z}_1, \vec{z}_2) \right ]^* ~ \Psi_A^0(\vec{x} ... \vec{y},
    \vec{z}'_1, \vec{z}'_2)
    \label{2.3.12}
 \ee
is the off-diagonal two-body density matrix. It should be pointed out that
the evaluation of both $n_{rel}^{N_1N_2}$ and $n_{CM}^{N_1N_2}$ requires the
knowledge of the ground-state wave function only. Till now, calculations of
the relative and $CM$ momentum distributions, obtained within many-body
approaches, have been reported only for $A = 3$ \cite{FAD} and $A = 4$
\cite{MAT88}.

\indent Before closing this subsection, we would like to mention that all the
calculations of $n(k)$ considered in this paper are based on the use of a
non-relativistic nuclear wave function. It is only recently that estimates of
the relativistic corrections to the nuclear Hamiltonian has been calculated in
case of light nuclei \cite{SCH93}; however, the effects of such corrections
on $n(k)$ turns out to be quite small up to $k \sim 4 \div 5 ~ fm^{-1}$, which
is just the limit considered in this paper.

\vspace{1cm}

{\large{\bf 2.4 The saturation of the momentum sum rule.}}

\vspace{0.5cm}

\indent 	In Ref. \cite{HE3} a relevant relationship between high momentum and
high removal energy components has been for the first time illustrated by
considering the partial momentum distribution $n_f(k)$, defined as follows
 \be
    n_f(k) \equiv 4 \pi ~ \int_{E_{min}}^{E_f} dE ~ P(k,E)
    \label{2.5.1}
 \ee
where the upper limit of integration $E_f$ can be varied from $E_{min}$ to
$\infty$. By definition, the partial momentum distribution $n_f(k)$
represents that part of $n(k)$ which is due to final ($A-1$)-nucleon states
with $E \leq E_f$. When $E_f \rightarrow \infty$, one gets $n_f(k) \rightarrow
n(k)$, and the momentum sum rule (Eq. (\ref{2.3.3})) is recovered. Thus, the
behaviour of $n_f(k)$ as a function of $E_f$ provides information on the
saturation of the momentum sum rule and the relevance of binding effects. The
calculation of $n_f(k)$ clearly requires the knowledge of the nucleon
Spectral Function and, therefore, till now has been performed only for $A
=3 $ and $A = \infty$. The results of \cite{HE3} for $^3He$ are reported in
Fig. 3(a) for various values of $E_f$. It can be seen that at low values of
$k$ ($ < 1.5~fm^{-1}$) the momentum sum rule is saturated already at values of
$E_f$ very close to $E_{min}$; this is not surprising, for at low momenta the
nucleon Spectral Function is dominated by its component $P_0(k,E)$, whose
strength is almost totally concentrated at low values of the removal energy
(cf. Eq. (\ref{2.2.12}) for $^3He$ and also Eq. (\ref{2.2.8.nm}) in case of
infinite nuclear matter). On the contrary, at high values of $k$ ($>
1.5~fm^{-1}$) the momentum sum rule is saturated only when high values of
$E_f$ are considered in Eq. (\ref{2.5.1}); one finds that the higher the
value of $k$ the higher the value of $E_f$ needed to saturate the momentum
sum rule. These features of the saturation of the momentum sum rule at high
momenta are due to the fact that the correlated part $P_1(k,E)$ of the
Spectral Function is spread over a wide range of values of $E$ for a given
$k$, in agreement with the theoretical predictions of many-body approaches
(see Ref. \cite{CSFS91}) as well as with the experimental data on the
exclusive processes $^3He(e,e'p)X$ \cite{MAR88} and $^4He(e,e'p)X$
\cite{MAG94}, in which the removal energy and momentum dependences of the
nucleon emission roughly follow the kinematics of the emission from a
two-nucleon system. Recently, such an important feature of the saturation of
the momentum sum rule at high momenta has been also demonstrated for infinite
nuclear matter in Ref. \cite{BFF}, as exhibited in Fig. 3(b).

\vspace{1cm}

{\Large{\bf 3. The convolution model for the nucleon Spectral Function.}}

\vspace{0.5cm}

\indent To develope our model Spectral Function, we make use of the two main
results emerging from the analysis of existing calculations of $n(k)$ and
$n_f(k)$ presented in the previous Section, namely: i) the "deuteron-like"
tail of $n(k)$, i.e., the observation that, apart from an overall scale
factor, the behaviour of $n(k)$ at high momenta ($k > 2 ~ fm^{-1}$) is almost
independent of the atomic weight $A$; ii) the saturation of the momentum sum
rule at high momenta, which clearly indicates the strong link between high
momentum and high removal energy components of the nuclear wave function,
which are both generated by $NN$ correlations. Both features should reflect
some local properties of the $NN$ wave function in the nuclear medium at short
internucleon separations.

\vspace{1cm}

{\large{\bf 3.1 The two-nucleon correlation model.}}

\vspace{0.5cm}

\indent The analysis of the momentum distributions and the saturation
of the momentum sum rule presented in Section 2, show that high momentum
and high removal energy components in nuclei are generated by $NN$
correlations resembling the ones acting in a deuteron, with the many body
aspects appearing through the constant $C^A$ (cf. eq.(29)) and the rich
spectrum of removal energy values of the spectral function for an $A>2$
nucleus. A first microscopic model of $NN$ correlations generating both high
momentum and high removal energy components has been proposed in \cite{FS},
where, by analyzing the perturbative expansion of  the $NN$ interaction
and the nucleon momentum distribution for $NN$ potentials decreasing at large
$k$ as powers of $k$, it has been argued  that the nucleon Spectral
Function at high values of both $k$ and $E$ should be governed by
ground-state configurations in which the high-momentum $\vec{k}_1 \equiv
\vec{k}$ of a nucleon is almost entirely balanced by the momentum $\vec{k}_2
\simeq - \vec{k}$ of another nucleon, with the remaining ($A-2$) nucleons
acting as a spectator with momentum $\vec{k}_{A-2} \simeq 0$
\footnote[2]{Configurations corresponding to high values of $|\vec{k}_{A-2}|$
should be ascribed to three-nucleon correlations; indeed, high values of
$|\vec{k}_{A-2}|$ can be due to ground-state configurations with a third
"hard" nucleon, whose momentum balances the $CM$ one of particles $1$ and
$2$.}. When the momentum and the intrinsic excitation energy of the ($A-2$)
system are totally disregarded, the energy conservation would require that
 \be
    E_{A-1}^* + E_{A-1}^R \simeq {k^2 \over 2M}
    \label{3.1.1}
 \ee
where $E_{A-1}^R = k^2 / 2 (A-1) M$ is the recoil energy of the
($A-1$)-nucleon system; thus, the intrinsic excitation of the ($A-1$) system
would be
 \be
    E_{A-1}^* = {A - 2 \over A-1} ~ {k^2 \over 2M}
    \label{3.1.2}
 \ee
Within such a picture, the nucleon Spectral Function $P_1(k,E)$ has the
following form
 \be
    P_1(k,E) = {1 \over 4 \pi} n_1(k) ~ \delta[E - E_1^{(2NC)}(k)]
    \label{3.1.3}
 \ee
with
 \be
    E_1^{(2NC)}(k) = E_{thr}^{(2)} ~ + ~ {A - 2 \over A-1} ~ {k^2 \over 2M}
    \label{3.1.4}
 \ee
where $E_{thr}^{(2)} \equiv |E_A| - |E_{A-2}|$ is the two-nucleon break-up
threshold.

\indent Let us denote by $E_1^{peak}(k)$ the value of the removal energy at
which, for a given momentum $k$, the Spectral Function $P_1(k,E)$ has its
maximum, and by $<E(k)>_1$ the mean removal energy for a given $k$, viz.
 \be
    <E(k)>_1 \equiv {4 \pi \over n_1(k)} \int_{E_{min}}^{\infty} dE ~ E ~
    P_1(k,E)
    \label{3.1.5}
 \ee
Using the Spectral Function given by Eq. (\ref{3.1.3}) one gets
 \be
    <E(k)>_1 ~ = ~ E_1^{peak}(k) ~ = ~ E_1^{(2NC)}(k)
    \label{3.1.6}
 \ee
In what follows, the model given by Eq. (\ref{3.1.3}) will be referred to as
the two-nucleon correlation ($2NC$) model. At high values of $k$ and $E$, the
nucleon Spectral Function calculated in $^3He$ \cite{CPS80} and nuclear
matter \cite{BFF} exhibits, indeed, for fixed values of $k$, broad peaks
in $E$, whose width increases with $k$. This is illustrated in Fig. 4, where
the Spectral Function in $^3He$ and infinite nuclear matter are shown for
$k > 2  fm^{-1}$ and $E > 50 MeV$ . In order to emphasize the high-momentum
part of $P_1(k,E)$ due to $NN$ correlations, the quantity
$k^2 \cdot P_1(k,E)$ has been plotted; it can
indeed be seen that for fixed values of $k$,  a peak-shaped behaviour is
exhibited, which can be characterized by three relevant quantities, viz. the
peak position $E_1^{peak}(k)$, the mean removal energy $<E(k)>_1$ defined by
Eq. (\ref{3.1.5}) and the Full Width @ Half Maximum (FWHM). In  Ref.
\cite{CSFS91} a quantitative comparison between the $2NC$ model and the
many-body Spectral Function has been presented. The results are reported in
Fig. 5, which shows an impressive agreement between the value of $<E(k)>_1$
calculated with the many-body Spectral Function and the predictions of the
$2NC$ model. However, it can also be seen that the $2NC$ model cannot predict
the difference between $<E(k)>_1$ and $E_1^{peak}(k)$ obtained within
many-body calculations. Moreover, the $2NC$ model, by definition, cannot
provide values of the Spectral Function for $E \neq E_1^{2NC}(k)$.

\vspace{1cm}

{\large{\bf 3.2 The extended two-nucleon correlation model.}}

\vspace{0.5cm}

\indent In Ref. \cite{CSFS91} the $NN$ correlation mechanism, which produces a
nonvanishing Spectral Function for $E \neq E_1(k)$, has been found to be the
motion of the $CM$ of the correlated $NN$ pair, and an expression of
$P_1(k,E)$ in terms of a convolution integral of the relative and $CM$
momentum distributions of a correlated pair has been derived in details for
the case of $^3He$. In this Section we show that a convolution formula can be
obtained for any value of the mass number $A$, and that such a convolution
formula follows from the high $k-E$ limit of the Spectral Function. We will
obtain our model of the Spectral Function of a nucleon $N_1$
($N_1 = n,p$) starting from the definition (\ref{2.1.4}). We are interested
in 2p-2h (1p-2h) configurations of $\Psi_A^0$ ($\Psi_{A-1}^{f_{A-1}}$)
appearing in the overlap integral (\ref{2.1.3}). The following Jacobi
coordinates and conjugate momenta referring to particles $1$ and $2$ (with
rest mass $M$) in the continuum, and "particle" $3$ (the ($A-2$)-nucleon
system) with rest mass $(A-2)M$, will be used
 \be
    \vec{x} = \vec{r}_1 - \vec{r}_2 ~~~~~~~~~~~~~~
    \vec{k}_x & = & {\vec{k}_1 - \vec{k}_2 \over 2} \nonumber \\
    \vec{y} = \vec{r}_3 - {\vec{r}_1 + \vec{r}_2 \over 2} ~~~~~~~~
    \vec{k}_y & = & {2 \vec{k}_3 - (A-2) (\vec{k}_1 + \vec{k}_2) \over A}
    \label{3.2.3}
 \ee
In what follows the sets of coordinates $\{ \vec{r}_i \}$ and conjugate
momenta $\{ \vec{k}_i \}$ are measured from the $CM$ of the nucleus (i.e.,
they satisfy the relations $\sum_{i=1}^A \vec{r}_i = 0$ and $\sum_{i=1}^A
\vec{k}_i = 0$), so that one has:
 \be
   \vec{k}_x & = & {\vec{k}_1 - \vec{k}_2 \over 2} \equiv \vec{k}_{rel}
   \nonumber \\
   \vec{k}_y & = & \vec{k}_3 = - (\vec{k}_1 + \vec{k}_2) \equiv -
   \vec{k}_{CM}
   \label{3.2.3bis}
\ee
where $\vec{k}_{rel}$ and $\vec{k}_{CM}$ are the relative and $CM$ momentum of
the correlated $NN$ pair, respectively. In terms of these variables the
ground-state wave function can be written in the following general form
 \be
    \Psi_A^0(\{ \vec{r}_i \}_A) = \hat{\cal{A}}_A \left \{ \sum_{n m f_{A-2}}
    a_{n m f_{A-2}} ~ \left [\Phi_n(\vec{x}) \otimes \chi_m(\vec{y}) \right ]
    \otimes \Psi_{A-2}^{f_{A-2}}( \{ \vec{r}_i \}_{A-2} ) \right \}
    \label{3.2.4}
 \ee
where $\hat{\cal{A}}$ is a proper antisymmetrization operator; $\otimes$ is a
short-hand notation for the standard Clebsh-Gordan coupling of orbital and
spin angular momenta; $\{ \Phi_n(\vec{x}) \}$ ($\{ \chi_m(\vec{y}) \}$)
represents a complete set of states describing the relative ($CM$) motion of
the pair ($1, 2$) and  $\{ \Psi_{A-2}^{f_{A-2}}( \{ \vec{r}_i \}_{A-2} ) \}$
the complete set of states describing the ($A-2$) system. The ansatz
(\ref{3.2.4}) is the exact one needed to calculate the Spectral Function
corresponding to states in which two particles are in the continuum. Our aim,
however, is to describe the high momentum and high removal energy parts of
the Spectral Function. To this end, we adhere to the argument that such a part
of $P_1(k,E)$ is generated by ground-state configurations in which two
particles are strongly correlated, with the ($A-2$) particles simply creating
the mean field in which the correlated pair is moving. If we view such a
configuration in momentum space, we would say that we are dealing with a $NN$
correlated pair with a very high relative momentum ($|\vec{k}_x| > k_f \sim
1.5 ~ fm^{-1}$) and a low $CM$ momentum ($|\vec{k}_y| < 1.5 ~ fm^{-1}$); these
assumptions allows one to safely describe the $CM$ motion wave function in
(\ref{3.2.5}) by a $0s$ state (which we denote by $m=0$), obtaining
 \be
    \Psi_A^0(\{ \vec{r}_i \}_A) \simeq \hat{\cal{A}} \left \{
    \chi_0(\vec{y}) \sum_{n f_{A-2}} a_{n 0 f_{A-2}} ~ \left [ \Phi_n(\vec{x})
    \otimes \Psi_{A-2}^{f_{A-2}}( \{ \vec{r}_i \}_{A-2} ) \right ] \right \}
    \label{3.2.5}
 \ee
To further elaborate on the structure of the wave function (\ref{3.2.5}),
we make use of the results illustrated in subsection 2.3, namely, since
the $CM$ of the pair involves only low-momentum components, the excitation
spectrum of the ($A-2$) system ($\{ f_{A-2} \}$) is mainly limited to the
ground-state and to the (low - lying) excited states corresponding to
configurations generated by the removal of two particles from different shell
model states of the target; thus, denoting such states by $f_{A-2} =
\bar{0}$, we have
 \be
    \Psi_A^0(\{ \vec{r}_i \}_A) \simeq \hat{\cal{A}} \left \{
    \chi_0(\vec{y}) ~ \left [ \Phi(\vec{x}) \otimes \Psi_{A-2}^{\bar{0}}( \{
    \vec{r}_i \}_{A-2} ) \right ] \right \}
    \label{3.2.6}
 \ee
where
 \be
    \Phi(\vec{x}) = \sum_n a_{n 0 \bar{0}} ~ \Phi_n(\vec{x})
    \label{3.2.7}
 \ee
describes the relative motion of the correlated pair in the nuclear medium,
which at small $|\vec{x}|$ can be considered independent of the particular
shells from which the two nucleons of the pair are removed by the effects of
short-range $NN$ correlations. Let us reiterate that Eq. (\ref{3.2.6}) is our
basic starting point for obtaining the Spectral Function. The basic
assumption underlying Eq. (\ref{3.2.6}) is that high-momentum components in
nuclei are due to strong correlations between two nucleons, whose $CM$ momentum
is a low one (which, in turns, means that $|\vec{x}| << |\vec{y}|$); once
such an assumption is made, it follows that the ($A-2$) system is in its
ground - state; the nucleons belonging to the ($A-2$) "spectator" system may
well be strongly correlated between themselves, but the basic assumption is
that they are almost independent from the "active" correlated pair which
"feels" the ($A-2$) system only through the low - momentum $CM$ motion. An
important feature of our $2NC$ model is that high values of the excitation
energy of the residual ($A-1$) system are almost totally due to high values
of the kinetic energy of the relative motion of the correlated nucleon with
the "spectator" ($A-2$) system, i.e., they are not generated by high values
of the excitation energy of the ($A-2$) system. To sum up, we assume that the
ansatz (\ref{3.2.6})  can describe the real configurations leading to the
high momentum and high removal energy components of the Spectral Function.

\indent Let us now discuss the final states $\Psi_{A-1}^{f_{A-1}}$ of the
residual ($A-1$)-nucleon system. The wave function $\Psi_{A-1}^{f_{A-1}}$ is
approximated as \be
    \Psi_{A-1}^{f_{A-1}}(\{ \vec{r}_i \}_{A-1}) = \hat{\cal{A}}_{A-1} ~
    \left \{ e^{i \vec{k}_2 \cdot \vec{r}_2} \otimes
    \Psi_{A-2}^{\tilde{f}_{A-2}}( \{ \vec{r}_i \}_{A-2} ) \right \}
    \label{3.2.8}
 \ee
where the distortion effects arising from the rescattering of particle $2$
with "particle $3$" have been neglected. The form (\ref{3.2.9}) is the
natural extension to the final states of the residual system of the basic
assumption underlying Eq. (\ref{3.2.5}) for the initial state. As a matter
of fact, after the removal of particle $1$ from a correlated pair, the
correlated particle $2$ is emitted because of recoil and "feels" the mean
field produced by the ($A-2$) system only in the low-momentum part of its
final amplitude. In other words, the final state interaction of particle $2$
is suppressed by the fact that the configurations relevant for the high
$k$-$E$ part of the nucleon Spectral Function are those in which $|\vec{x}|
<< |\vec{y}|$. It should be pointed out that in case of $^3He$ explicit
many-body calculations  \cite{CPS78} of the nucleon Spectral Function clearly
show that the effects of the interaction in the residual system are relevant
only for the low removal-energy part of $P_1(k,E)$, whereas they are
negligible for the high $k$-$E$ component which we are interested in.

\indent Placing Eqs. (\ref{3.2.6}) and (\ref{3.2.8}) into Eq. (\ref{2.1.3})
and omitting spin indeces for sake of simplicity, we get for the Fourier
transform of the overlap integral the following expression
 \be
    \int d\vec{z} e^{i \vec{k} \cdot \vec{z}} ~ G_{f0}(\vec{z}) & \propto &
    \int d\vec{x} d\vec{y} ~ e^{i \vec{k}_x \cdot \vec{x} + i \vec{k}_y \cdot
    \vec{y}} ~ \chi_0(\vec{y}) ~ \Phi(\vec{x}) = \tilde{\chi}_0(\vec{k}_y) ~
    \tilde{\Phi}(\vec{k}_x)
    \label{3.2.9}
 \ee
where $\tilde{\Phi}(\vec{k}_x)$ ($\tilde{\chi}_0(\vec{k}_y)$) is the
amplitude for the relative ($CM$) motion of the correlated pair in momentum
space. Within our model the energy conservation appearing in Eq.
(\ref{2.1.4}) reads as follows
 \be
    E - E_{A-1}^f + E_A & = & E +|E_{A-2}| - {|\vec{t}|^2 \over 2 \mu} -
    |E_A| \nonumber \\
    & = & E - \left ( E_{thr}^{(2)} + {|\vec{t}|^2 \over 2 \mu} \right )
    \label{3.2.10}
 \ee
where
 \be
    {|\vec{t}|^2 \over 2 \mu} \equiv {(A-1) \over 2 M (A-2)} \left [ {(A-2)
    \vec{k}_2 - \vec{k}_3 \over A-1} \right ]^2 = {(A-2) \over 2 M (A-1)}
    \left [ \vec{k} + {(A-1) \vec{k}_3 \over A-2} \right ]^2
    \label{3.2.11}
 \ee
is the energy of the relative motion of particle $2$ and $"3"$ ($A-2$), i.e.
the excitation energy of the ($A-1$) system. Thus, placing Eqs. (\ref{3.2.9})
- (\ref{3.2.11}) into Eq. (\ref{2.1.4}) the following convolution formula is
obtained for any value of $A$
 \be
    P_1^{(N_1)}(k,E) & = & {\cal{N}} ~ \sum_{N_2 = n,p} ~ \int d \vec{k}_3
    ~ \delta \left [ E - E_{thr}^{(2)} - {(A-2) \over 2M (A-1)} \left (
    \vec{k} + {(A-1) \vec{k}_3 \over (A-2)} \right ) ^2 \right ] \nonumber \\
    & \cdot & n_{rel}^{N_1N_2} (| \vec{k} + {\vec{k}_3 \over 2} |) ~
    n_{CM}^{N_1N_2} (| \vec{k}_3 |)
    \label{3.2.12}
 \ee
In Eq. (\ref{3.2.12}) the value of ${\cal{N}}$ can be chosen in order to
reproduce the correlated part of the nucleon momentum distribution $n_1(k) =
n(k) - n_0(k)$ ($\simeq n(k)$ for $k > 1.5 ~ fm^{-1}$), viz.
 \be
    \int_{E_{thr}^{(2)}}^{\infty} dE ~ P_1^{(N_1)}(k, E) = n_1(k)
    \label{3.2.13}
 \ee
{}From Eq. (\ref{3.2.12}) it can be seen that the $2NC$ model can be recovered
by placing $n_{CM}^{N_1N_2} (\vec{k}_3) = \delta(\vec{k}_3)$, i.e. the
spectator nucleon at rest. When the motion of the latter and the link between
$\vec{k}$, $\vec{k}_3$ and $E$ ares considered, a nucleon Spectral Function in
the whole range of variation of $E$ ($E_{thr}^{(2)} \leq E < \infty$) is
generated. Moreover, the removal energy dependence of $P_1^{(N_1)}(k,E)$ is
governed by the behaviour of $n_{CM}^{N_1N_2}$ and $n_{rel}^{N_1N_2}$, whose
calculation, unlike the case of the Spectral Function itself, requires the
knowledge of the ground-state wave function only.

\indent A further integration in Eq. (\ref{3.2.12}) over the angular variables
of $\vec{k}_3$ yields
 \be
    P_1^{(N_1)}(k,E) = {2 \pi M \over k} {\cal{N}} ~ \sum_{N_2 = n,p} ~
    \int_{k_3^-}^{k_3^+} d k_3 ~ k_3 ~ n_{rel}^{N_1N_2} (k_x^*) ~
    n_{CM}^{N_1N_2} (k_3)
    \label{3.2.14}
 \ee
where $k_3 \equiv |\vec{k}_3|$ and
 \be
   & & k_3^{\pm} = {A-2 \over A-1} |k \pm k_0| \nonumber \\
   & & k_0 = \sqrt{2M {A-1 \over A-2} [E - E_{thr}^{(2)}]} \nonumber \\
   & & k_x^* = \sqrt{{A k^2 + (A-2) k_0^2 \over 2(A-1)} - {A k_3^2 \over
               4(A-2)}}
   \label{3.2.15}
 \ee
It can be easily seen that Eq. (\ref{3.2.14}) reduces for $A=3$ to the
convolution formula given by Eq. (18) of Ref. \cite{CSFS91}. As already
pointed out, in Eq. (\ref{3.2.14}) $n_{rel}^{N_1N_2}$ refers to a proper spin
and isospin combination of a $N_1N_2$ pair in the continuum, and,
correspondingly, $n_{CM}^{N_1N_2}$ represents the momentum distribution of a
$N_1N_2$ pair with respect to the ($A-2$) system in its ground state. The
"three-body" configuration underlying Eq. (\ref{3.2.14}) is such that two
correlated particles are very close, whereas the ($A-2$) core is far from
their $CM$. Therefore, in Eq. (\ref{3.2.14}) the relevant contribution has to
be provided by the low-momentum part of $n_{CM}^{N_1N_2}$ and by the
high-momentum one of $n_{rel}^{N_1N_2}$. Such a ground-state configuration is
automatically generated by the use of the $CM$ momentum distribution in the
$\ell = 0$ (mean field) state, which, for $k > 2 ~ fm^{-1}$, does not include
the high-momentum components generated by the short-range and tensor
correlations. We would like to stress here that, according to our
assumptions, Eq. (\ref{3.2.14}) is expected to correctly describe the nucleon
Spectral Function only for high values of $k$ and $E$, when the ($A-2$)
system is in its ground state; in terms of $n_{rel}^{N_1N_2}$ and
$n_{CM}^{N_1N_2}$ it means that only high (low) values of $k_x^*$ ($k_3$)
have to be considerd in Eq. (\ref{3.2.14}).

\vspace{1cm}

{\Large{\bf 4. The model Spectral Function for few-nucleon systems and
complex nuclei.}}

\vspace{0.5cm}

\indent In this Section our model Spectral Function will be presented for $3
\leq A < \infty$. For the three-nucleon system and infinite nuclear matter our
model Spectral Function will be compared with the one obtained from
many-body calculations. According to Eq. (\ref{3.2.14}), the basic
ingredients to calculate the Spectral Function are the relative and $CM$
momentum distributions of $NN$ pair in the nucleus. These quantities have
been calculated for $A = 3, 4$ and their behaviour will be discussed togheter
with model distributions for complex nuclei.

\vspace{1cm}

{\large{\bf 4.1 The relative and $CM$ momentum distributions.}}

\vspace{0.5cm}

\indent The quantities (\ref{2.3.10}) and (\ref{2.3.10bis}) pertaining to a
neutron-proton and neutron (proton) - neutron (proton) pairs in $^3He$ and
$^4He$, calculated in Refs. \cite{CPS80} and \cite{MAT88} using the Reid Soft
Core interaction \cite{RSC}, are presented in Fig. 6, where they are
compared with a rescaled deuteron momentum distribution for the high-momentum
part of the relative distributions and with a simple Gaussian parametrization
for the low-momentum part of the $CM$ distributions (see also Eq.
(\ref{3.2.3bis}) which relates the relative and $CM$ momenta ($\vec{k}_{rel}$
and $\vec{k}_{CM}$) with the Jacobian momenta $\vec{k}_x$ and $\vec{k}_3$
appearing in Eq. (\ref{3.2.14})). It can be seen that:  i) the high-momentum
part of the relative distribution can be very accurately explained by a
rescaled deuteron momentum distribution, as suggested by several
investigations; therefore, the following effective relative momentum
distribution will be used in Eq. (\ref{3.2.14})
 \be
   n_{rel}^{eff}(k_{rel}) = C^A ~ n_D(k_{rel})
   \label{4.1.1}
\ee
where $C^A$ is the same constant appearing in Eq. (\ref{2.3.9})
\footnote[3]{Note that by using  (\ref{4.1.1}) in (\ref{3.2.12}) one gets,
due to (\ref{2.3.9}), that  ${\cal{N}} \simeq 1$ for $k > 2 ~ fm^{-1}$}; ii)
the low-momentum part of $n_{CM}^{N_1N_2}$ can be fairly well reproduced by a
Gaussian distribution, in agreement with our assumption that the $CM$ moves
in a $0s$ state; thus, for complex nuclei we will use in Eq. (\ref{3.2.14})
the following effective $CM$ momentum distribution
 \be
    n_{CM}^{eff}(k_{CM}) = ({\pi \over \alpha_{CM}})^{{3 \over 2}} ~
    e^{ - \alpha_{CM} k_{CM}^2}
    \label{4.1.2}
 \ee
with the parameter $\alpha_{CM}$ determined as follows. Let us start from
the trivial relation $< (\sum_{i=1}^A \vec{k}_i)^2> = 0$, where the
expectation value is performed with respect to the intrinsic wave function
$\Psi_A^0$; such a relation implies $A <k^2> + A (A-1) <\vec{k}_1 \cdot
\vec{k}_2> = 0$, where $<k^2>$ is the mean value of the squared single nucleon
momentum; then $<k_{CM}^2> = 2 (A-2) <k^2> / (A-1)$. Since, according to our
model, the distribution (\ref{4.1.2}) should not include high-momentum
components generated by short-range and tensor correlations, the value of
$<k^2>$ is taken to be equal to the one obtained within the mean-field
approach. Therefore, $\alpha_{CM}$ appearing in Eq. (\ref{4.1.2}) is given by
 \be
    \alpha_{CM} = {3 \over 2 <k_{CM}^2>} = {3 (A-1) \over 4 (A-2)} {1 \over 2M
    <T>^{(SM)}}
    \label{4.1.3}
 \ee
where $<T>^{(SM)}$ is the expectation value of the (single) nucleon kinetic
energy calculated within the mean-field approach. The value of $\alpha_{CM}$
determined by Eq. (\ref{4.1.3}) for various nuclei as well as the value of
the parameter $C^A$ Appearing in Eq. (\ref{4.1.1}) are listed in Table 1.

\vspace{1cm}

{\large{\bf 4.2 The nucleon Spectral Function of $^3He$ and nuclear matter.}}

\vspace{0.5cm}

\indent Here, the quantity $P_1(k,E)$ for $^3He$ and infinite nuclear matter
calculated from Eq. (\ref{3.2.14}) will be presented and compared with the
same quantity calculated from many-body approaches already discussed and
shown in Fig. 4. First of all, let us show the comparison between our model
predictions for the peak position $E_1^{peak}(k)$, the mean removal energy
$<E(k)>_1$ and the FWHM, with the predictions from many-body calculations. An
inspection at Eq. (\ref{3.2.14}) shows qualitatively the following relevant
features of the Spectral Function: i) if $n_{rel}$ and $n_{CM}$ are described
within an independent particle model (e.g., Gaussians with the same lenght
parameter), Eq. (\ref{3.2.14}) predicts a maximum of the Spectral Function
close to $E = E_{thr}^{(2)}$ with a monotonic decrease with $E$; ii) because
of the $k_3$ and $k_0$ dependence of the argument $k_x^*$ of $n_{rel}$, the
Spectral Function exhibits a peak-shaped behaviour with the peak position
located at a value lower than the one predicted by the $2NC$ model (see Eq.
(\ref{3.1.4})); iii) the shift of the peak position with respect to the
$2NC$ model, as well as the shape of the Spectral Function near the peak, are
mostly governed by the high $k_0$ dependence of $n_{rel}(k_x^*)$. If the
latter is chosen to be of the Gaussian form, the peak position and the FWHM
can be expressed as a series expansion in terms of the parameter
 \be
    \gamma \equiv {A-1 \over 2 (A-2)} ~ <k_{CM}^2> / <k_{rel}^2>
    \label{4.2.1}
 \ee
where $<k_{rel}^2>$ and $<k_{CM}^2>$ are the mean-square momenta associated
to the high and the low-momentum parts of $n_{rel}$ and $n_{CM}$,
respectively. One has
 \be
    E_1^{peak}(k) \simeq E_{thr}^{(2)} + {A-2 \over A-1} ~ {k^2 \over 2M} ~
    [1 - 2 \gamma] ~ + ~ O(\gamma^2)
    \label{4.2.2}
 \ee
 \be
    FWHM(k) \simeq \sqrt{{<k_{CM}^2> ~ 8 ln2 \over 3}} ~ (1 - \gamma) ~ {k
    \over M} ~ + ~ O(\gamma^2)
    \label{4.2.3}
 \ee
Equations (\ref{4.2.2}) and (\ref{4.2.3}) show again that in the limit of
a static spectator ($A-2$) system (i.e., $<k_{CM}^2> \rightarrow 0$) the $2NC$
model is recovered (i.e., $E_1^{peak}(k) = E_{thr}^{(2)} + (A-2) k^2 /
2(A-1)M$ and $FWHM = 0$). The motion of the spectator system, coupled,
through energy and momentum conservation, to the relative motion of the
correlated pair (cf. Eq. (\ref{3.2.14})), produces both a shift (by a
percentage aumont of the order of $\simeq 2 \gamma$) of the peak position from
the value predicted by the $2NC$ model, as well as the removal energy
dependence of the Spectral Function for $E \neq E_1^{peak}(k)$. The validity
of the above expressions relies on the smallness of the parameter $\gamma$.
{}From many-body calculations \cite{FAD}(b), \cite{CPS80} and \cite{BFF} one
gets $<k_{rel}^2> \sim 6 ~ fm^{-2}$ ($\sim 7 ~ fm^{-2}$) and $<k_{CM}^2> \sim
0.5 ~ fm^{-2}$  ($\sim 2 ~ fm^{-2}$) for $^3He$ (nuclear matter), so that
$\gamma_{^3He} \sim 0.08$ ($\gamma_{NM} \sim 0.14$). In Figs. 8 and 9
$E_1^{peak}(k)$ and FWHM calculated using Eqs. (\ref{4.2.2}) and
(\ref{4.2.3}), as well as using the full Spectral Function, are compared with
the corresponding quantities predicted by many-body Spectral Functions shown
in Fig. 4. It can be seen that: i) unlike the $2NC$ model in which $<E(k)>_1
= E_1^{peak}(k)$, in disagreement with the results of theoretical
calculations, Eq. (\ref{3.2.14}) is able to correctly predict the relation
$<E(k)>_1 > E_1^{peak}(k)$ (cf. Fig. 7); ii) the values of the FWHM, which is
obviously zero in the $2NC$ model, is correctly predicted by Eq.
(\ref{3.2.14}). It turns out also that from the results presented in Fig. 8,
the linear dependence of the FWHM upon $k$ (see Eq. (\ref{4.2.3})) provides a
satisfactory reproduction of the average value of the FWHM up to large value
of $k$; at the same time, it appears that the calculations with the many-body
Spectral Function also give rise to terms quadratic in $k$, which are
reproduced by evaluating Eq. (\ref{3.2.14}) with realistic momentum
distributions. Eventually, in Fig. 9  we present a direct comparison of
our model Spectral Function with the many-body ones. It can be seen that the
whole shape of $P_1(k,E)$ is satisfactorily reproduced in a wide range of
values of $E$ around the peak. It can be noticed that the range of $k$ and
$E$ for which our model can be applied is wider in nuclear matter than in
$^3He$. This is a typical $A$-dependent effect, since it is due to the $A$
dependence of the FWHM (cf. (\ref{4.2.3})). In Fig.10 we provide a three-
dimensional plot of the Spectral Functions; there it can be seen that,
whereas for nuclear matter the agreement between the model and many-body
Spectral Functions is a satisfactory one in the whole range of momenta and
energies considered, in ${^3He}$ an appreciable disagreement in the low $k$ -
high $E$ region can be observed.  Such a disagreement, which,
we reiterate, appears in $^3{He}$ but not in nuclear matter, can be understood
as follows: our model includes two-nucleon correlations only, i.e. it cannot
account for three and more nucleon correlations, which correspond to
configurations in which more than two particles get close by. Since  the
"exact"  $^3He$  Spectral Function  includes three-nucleon correlations,
and since these mainly affect very low and very high energy tails,
the reason for the disagreement in $^3He$ is clear: it is due to the
absence of three-nucleon correlations in our model. On the other hand side,
in the case  of nuclear matter, both the many-body \cite{BFF} and our model
Spectral Functions do not contain three-nucleon correlations and therefore
they agree even in the low $k$ - high $E$ region. Thus, we can conclude that
our model Spectral Function describes correctly the $k-E$ dependence of the
real Spectral Function generated by two-nucleon correlations in the range
$k > 2 ~ fm^{-1}$ and $E_L < E < E_1^{peak}(k) + FWHM(k)$, where
$E_1^{peak}(k)$ and the FWHM can be safely estimated by Eqs. (\ref{4.2.2})
and (\ref{4.2.3}). From Fig. 9 the value of $E_L$ can be estimated to be
$\sim 20 \div 30 ~ MeV$ in $^3He$ and $\sim 40 \div 50 ~ MeV$ in nuclear
matter. Moreover, we would like to point out that, using Eq. (\ref{3.2.15}),
the lower limit of integration $k_3^-$ in Eq. (\ref{3.2.14}), calculated
at $E = E_1^{peak}(k) + FWHM(k)$, is less than $0.5$ ($1$) $fm^{-1}$ in $^3He$
(nuclear matter) and the corresponding value of the relative momentum
$k_x^*$ is greater than $2 ~ fm^{-1}$.

\vspace{1cm}

{\large{\bf 4.3 The nucleon Spectral Function of $^4He$ and multiparticle
final states.}}

\vspace{0.5cm}

\indent To describe the Spectral Function for $A = 4$ our model has to be
implemented to take into account the so-called four-body channel, i.e.,
the configurations of the ($A-1$) system with $3$ particles in the
continuum. The reason is as follows. The basic idea, leading to the
convolution integral (\ref{3.2.12}), was the assumption that 2p - 2h
ground-state correlations (cf. Eq. (\ref{3.2.5})) leading to 1p - 2h states of
the residual ($A-1$) system (cf. Eq. (\ref{3.2.8})), are due, at high values
of $k$ and $E$, to configurations in which the high momentum of particle $1$
is almost entirely balanced by the one of particle $2$ with the "third"
spectator particle (the ($A-2$) system) far apart from the two correlated
particles. The excitation energy of the ($A-1$) system corresponding to such
a configuration is mainly due (particularly around the peak where $\vec{k}_2
\simeq - \vec{k}$ and $|\vec{k}_3| \sim 0$) to the kinetic energy of the
relative motion of particles $2$ and $3$, with the latter (the ($A-2$)
system) being, as previously explained, in low-lying excited states. When
$E > E_1^{peak}(k)$ one has $\vec{k}_2 \neq - \vec{k}$ and, in particular,
one can have $|\vec{k}_2| < |\vec{k}|$, if high values of the excitation
energy of the ($A-2$) system are allowed. Such a mechanism is absent in our
model, which effectively takes into account only 2p - 2h ground-state
correlations and, as already pointed out, the agreement we found with the
many-body calculations of nuclear matter makes us confident of the
correctness of our approach. To sum up, in a complex nucleus the removal
energy behaviour of the Spectral Function at high values of $k$ is determined
only by the relative motion of particles $2$ and $3$. In $^4He$ the situation
appears to be rather different, for the ($A-2$) system is the weakly bound
deuteron. Thus, for fixed values of $E$ even a small difference between
$|\vec{k}_2|$ and $|\vec{k}|$ at the peak can originate an intrinsic
excitation of the residual ($A-2$) system, which is sufficient to break-up
the deuteron giving rise to the four-body channel. In other words, our model
represented by Eq. (\ref{3.2.12}) with the hard part of $n_{rel}$, cannot be
applied to the calculation of the two-nucleon emission channel, for
considering $n_{rel}^{hard}$ will unavoidably lead to a break-up of the
residual deuteron. Moreover, it should be reminded that only in the case of a
neutron-proton pair (with total isospin equal to $0$) the residual system can
be a deuteron, whereas for other types of correlated $NN$ pairs the ($A-2$)
system can be only in the continuum. Therefore we have to describe, within
our model, the four - body channel. This can easily be done by the following
steps: i) the summation over $f_{A-2}$ has to be kept in Eq. (\ref{3.2.5});
ii) the factorisation of the relative motion of the pair and the intrinsic
excitation of the ($A-2$) system is assumed to hold. The final result is
 \be
    P_1^{(N_1)}(k,E) & = & {\cal{N}}(k) ~ \sum_{N_2 = n,p} ~ \int d \vec{k}_3
    ~ n_{rel}^{N_1N_2} (| \vec{k} + {\vec{k}_3 \over 2} |) ~ n_{CM}^{N_1N_2}
    (|\vec{k}_3 |) \nonumber \\
    & \cdot & \int d\vec{t} ~ w(|\vec{t}|) ~ \delta \left [ E - {t^2 \over M}
    - {1 \over 3M} \left ( \vec{k} + {3 \vec{k}_3 \over 2} \right )^2 \right ]
    \label{4.3.1}
 \ee
where $w(|\vec{t}|)$ represents the probability distribution to have a
two-nucleon final state with excitation energy $t^2 / M$. We have evaluated
Eq. (\ref{4.3.1}) by replacing $w(|\vec{t}|)$ with the hard part of the
deuteron momentum distribution; the results are presented in Fig. 11 in
correspondance of different values of the upper limit of integration over
$|\vec{t}| = \sqrt{M E_{A-2}^*}$. In the same figure the results of the
contribution of the four-body channel calculated in Ref. \cite{MS91} are
also shown.

\vspace{1cm}

{\large{\bf 4.4 The nucleon Spectral Function for complex nuclei.}}

\vspace{0.5cm}

\indent Given the success of our model for $^3He$, $^4He$ and nuclear matter
we are confident that it can safely be applied to complex nuclei.
In Fig.12 we present the Spectral Function  for $^{16}O$; the results for
other complex nuclei look very similar. In Fig. 13 we show the
saturation of the energy and momentum sum rules, i.e. the quantities
 \be
    S_f(E) = 4 \pi \int_0^{k_f} dk ~ k^2 ~ P(k,E)
    \label{4.4.1}
 \ee
and
 \be
    n_f(k) = 4 \pi \int_{E_{min}}^{E_f} dE P(k,E)
    \label{4.4.2}
 \ee
As far as the latter quantity is concerned, the trend found for $^3He$ and
nuclear matter (see subsection 2.4) is confirmed (see Fig. 3); as for the
former one, it can be seen that in order to saturate the energy sum rule,
momentum components larger than $2 ~ fm^{-1}$ are necessary.

\indent Another quantity of interest is the energy-weighted sum rule
\cite{KOL74}
 \be
    \varepsilon_A = {1 \over M} \left [ {A - 2 \over A -1} <T> - <E> \right ]
    \label{4.4.3}
 \ee
which relates the total binding energy per particle ($\varepsilon_A$) to the
mean
value of the nucleon removal energy
 \be
    <E> = \int d\vec{k} dE ~ E ~ P(k,E)
    \label{4.4.4}
 \ee
and of the nucleon kinetic energy
 \be
    <T> = \int_0^{\infty} dk ~ k^2 ~ {k^2 \over 2M} ~ n(k)
    \label{4.4.5}
 \ee
Using Eq. (\ref{2.2.8}) for $P_0(k,E)$ and our extended $2NC$ model for
$P_1(k,E)$ the mean removal energy has been calculated for several complex
nuclei. In Table 2 our results are reported and compared with those obtained
using in Eq. (\ref{4.4.3}) the experimental values of $\varepsilon_A$ and the
values of $<T>$ calculated using in Eq. (\ref{4.4.5}) the nucleon
momentum distribution $n(k)$ obtained within many-body approaches. It can be
seen that our model Spectral Function satisfies the energy-weighted sum rule
(\ref{4.4.3}).

\indent Recently \cite{SICK94} a model Spectral Function for complex nuclei,
based on the application of the Local Density Approximation (LDA) to the
correlated part $P_1(k,E)$, has been proposed. A direct comparison between
the two approaches cannot be performed, since the behaviour of the Spectral
Function for complex nuclei obtained in Ref. \cite{SICK94} was not provided.
On the other hand side, it has been shown \cite{SICK94} that the nucleon
momentum distribution $n(k)$, calculated for several finite nuclei by
applying the LDA to the nucleon momentum distribution in infinite nuclear
matter, compares very favourably with the results of many-body calculations.
However, such an agreement can hardly be considered, in our opinion, a merit
of the LDA; as a matter of fact, the high-momentum tail of $n(k)$ in finite
nuclei and infinite nuclear matter are already approximately equal, thanks
to the scaling property of $n(k)$ with $A$ (cf. Eq. (\ref{2.3.9})).
We should also stress that it is difficult to compare our
Spectral Function with the $LDA$ one on the basis of the calculation of
inclusive cross sections (cf. Section 5), for in the two approaches the
treatment of Final State Interaction ($FSI$) is different. Indeed, in our
approach \cite{CS94} the $FSI$ is such that the inclusive cross section is
very sensitive upon the details of the correlated part $P_1(k,E)$ of the
Spectral Function, whereas in \cite{SICK94} the inclusive nuclear response is
only weakly affected by $P_1(k,E)$, when the final state interaction of the
struck nucleon is considered as in Ref. \cite{PAND91}. To sum up, it would be
very interesting to know the $k$ and $E$ behaviours of the $LDA$ Spectral
Function and to compare them with the ones obtained within our model (see,
e.g., Fig. 12). In this respect we would like to stress  that our Spectral
Function provides a microscopic interpretation, in terms of
two-nucleon correlations, of why the high $k-E$ behaviour of the nuclear
matter Spectral Function is as it is, i.e. featuring peaks at high values of
$E$ located at $E \propto k^2 / 2M$ with a $FWHM$ given by Eq. (\ref{4.2.3}).

\vspace{1cm}

{\large{\bf 5. $NN$ correlations and inclusive quasi-elastic electron
 scattering at $x>1$}}

\vspace{0.5cm}

\indent The effects of $NN$ correlations on inclusive electron scattering
has been extensively investigated in Ref. \cite{CPS80} for $^3He$ and
\cite{CLS90,CPS91,CDL92} for complex nuclei using the plane wave Impulse
Approximation ($IA$). Calculations of Refs. \cite{CLS90,CDL92} have been
performed using an approximated Spectral Function, elaborated in Ref.
\cite{CSFS89}. The present Spectral Function has been used for the first
time in Ref. \cite{CS94}, where the $FSI$ has also been taken into account.
In \cite{CS94} the differential cross section for inclusive quasi-elastic
(QE) electron scattering off nuclei, $A(e,e')X$, has been evaluated at high
values of the squared four momentum transfer $Q^2$ ($> 1 ~ (GeV/c)^2$). In
particular, the kinematical regions corresponding to $x > 1 + k_F / M \simeq
1.3$ (where $k_F$ is the Fermi momentum and $x = Q^2 / 2M\nu$ the Bjorken
scaling variable) have been considered, for it is in such regions that the
nuclear response, evaluated within the $IA$, is sharply affected by the high
$k-E$ components of the nuclear wave function generated by $NN$ correlations
(cf. Ref. \cite{CLS90,CPS91}). However, it is also well known that the $IA$
sizably underestimates the inclusive cross section at $x > 1.3$ and this fact
is usually ascribed to the lack of any $FSI$ effect within the $IA$. In
\cite{CS94} the role played both by $NN$ correlations and $FSI$ has been
addressed and the first results of a calculation of the inclusive cross
section based upon a novel approach for evaluating $FSI$ have been presented.
Such an approach relies on a consistent treatment of $NN$ correlations both
in initial and final states, by extending the factorization hypothesis
(\ref{3.2.6}) to the final nuclear states \cite{CS94}.

\indent The differential cross section for the inclusive process $A(e,e')X$
can be written in the following form
 \be
  \sigma^{(A)} \equiv \frac{d^2\sigma} {dE_{e'}~d\Omega_{e'}} = \sigma_0^{(A)}
  + \sigma_1^{(A)}
  \label{5.1}
 \ee
where the indeces $0$ and $1$ have the same meaning as in Eq. (\ref{2.2.1}),
i.e., they distinguish the contributions resulting from different final
nuclear states, namely $\sigma_0^{(A)}$ describes the transition to the
ground and one-hole states of the (A-1)-nucleon system and $\sigma_1^{(A)}$
the transition to more complex highly excited configurations. In Ref.
\cite{CS94}, the calculation of $\sigma_0^{(A)}$ and $\sigma_1^{(A)}$ has
been carried out adopting different levels of sophistication for the
treatment of the final A-nucleon state, starting with the $IA$.

\indent The inclusive cross sections $\sigma_0^{(A)}$ and $\sigma_1^{(A)}$
within the $IA$ reads as follows \cite{CPS91}
 \be
    [\sigma_0^{(A)}]_{IA} = \sum_{N=1}^A ~ \int d\vec{k} ~ dE ~ \sigma_{eN}
    ~ P_0^{(N)} (k,E) ~ \delta \left [ \nu + k^0 - E_{\vec{k} + \vec{q}}
    \right ]
    \label{5.2}
 \ee
 \be
    [\sigma_1^{(A)}]_{IA} = \sum_{N=1}^A ~ \int d\vec{k} ~ dE ~ \sigma_{eN}
    ~ P_1^{(N)} (k,E) ~ \delta \left [ \nu + k^0 - E_{\vec{k} + \vec{q}}
    \right ]
    \label{5.3}
 \ee
where $\nu ~ (\vec{q})$ is the energy (three-momentum) transfer; $\vec{k}$
is the  momentum of the nucleon in the lab system before interaction and
$k^0 = M_A - \sqrt{(M_A + E - M)^2 + k^2}$ its off-shell energy; $E_{\vec{p}} =
\sqrt{M^2 + | \vec{p} | ^2}$; $\sigma_{eN}$ is the electron - (off-shell)
nucleon cross section. Using Eq. (\ref{3.2.12}) for $P_1(k,E)$, Eq.
(\ref{5.3}) can be written in the following form
 \be
    [\sigma_1^{(A)}]_{IA} = A \sigma_{Mott} \sum_{N_1 N_2 =n,p} ~ \int d
    \vec{k}_{CM} ~ n_{CM}^{N_1N_2}(\vec{k}_{CM}) ~ L^{\mu \nu} ~ [W_{\mu
    \nu}^{N_1 N_2}]_{IA}
    \label{5.4}
 \ee
where $L_{\mu \nu}$ is the (reduced) leptonic tensor and $[W_{\mu \nu}^{N_1
N_2}]_{IA}$ the hadronic tensor of a correlated pair, which can be written as
 \be
   [W_{\mu \nu}^{N_1 N_2}]_{IA} & = & {k_{CM}^0 \over 2M} \sum_{f_{12}^0}
   \sum_{\beta_{12}} ~ \left [ \langle \beta_{12} | j_{\mu}^{N_1} +
   j_{\mu}^{N_2} | f_{12}^0 \rangle \right ]^* ~ \sum_{\beta'_{12}} \left [
   \langle \beta'_{12} | j_{\nu}^{N_1} + j_{\nu}^{N_2} | f_{12}^0 \rangle
   \right ] \nonumber \\
   & \cdot & \delta \left [ \nu + k_{CM}^0 - \sqrt{ (M_2^{f_{12}^0})^2 +
   (\vec{k}_{CM} + \vec{q})^2} ~ \right ]
   \label{5.5}
 \ee
where $j_{\mu}^N$ is the nucleon current, $k_{CM}^0 = M_A - \sqrt{M_{A-2}^2 +
 | \vec{k}_{CM} | ^2}$, $| \beta_{12} \rangle$ is the relative wave
function of the correlated pair and $| f_{12}^0 \rangle$ its plane wave final
state. Eq. (\ref{5.4}) is based upon the assumption that final and initial
A-nucleon states factorize as follows: $| \Psi_A^f \rangle \sim \hat{A} ~
\left [ | f_{12}^0 \rangle | \vec{P}_{CM} \rangle | \Psi_{A-2}^f \rangle
\right ]$ and  $| \Psi_A^0 \rangle \sim \hat{A} ~ \left [ | \beta_{12} \rangle
| \chi_{12}^{CM} \rangle | \Psi_{A-2}^0 \rangle \right ]$ (cf. Eq.
(\ref{3.2.6})), where $| \chi_{12}^{CM} \rangle$ is the $CM$ wave function of a
correlated pair and $| \vec{P}_{CM} \rangle$ its plane wave final state (note
that, using Eq. (\ref{5.5}) in Eq. (\ref{5.4}), Eq. (\ref{5.3}) is recovered
in terms of the nucleon spectral function $P_1(k,E)$ given by Eq.
(\ref{3.2.12}) with $n_{rel}^{N_1 N_2}(\vec{k}_{rel}) = \sum_{\beta_{12}}
\left | \langle \beta_{12} | \vec{k}_{rel} \rangle \right | ^2$).  In Ref.
\cite{CS94} the $FSI$ has been evaluated by a novel approach based upon the
observation that the $FSI$ involving two-nucleons emitted because of
ground-state $NN$ correlations (\em two-nucleon rescattering \rm) should be
different from the $FSI$ involving the outgoing nucleon knocked-out from shell
model states (\em single nucleon rescattering \rm).

\indent Within the spirit of our model, at high values of $k$ and $E$ the
absorption of the virtual photon by a correlated $NN$ pair, which at $x > 1.3$
is the dominant mechanism in the $IA$, is expected to resemble the one in the
deuteron; if so, the deuteron-like picture of the initial state should be
extended also to the final state by allowing the two nucleons to elastically
rescatter. An important difference with respect to the case of the deuteron is
that a correlated $NN$ pair in a nucleus is bound and moves in the field
created by the other nucleons. The basic step of our approach is the
replacement of the $IA$ hadronic tensor $[W_{\mu \nu}^{N_1N_2} ]_{IA}$ by the
interacting one $W_{\mu \nu}^{N_1 N_2}$, which is nothing but Eq. (\ref{5.5})
with the plane wave state $| f_{12}^0 \rangle$ replaced by the exact $NN$
scattering wave function $| f_{12} \rangle$ (note that the two-nucleon
rescattering process cannot be expressed in terms of a spectral function). It
can be seen from Eq. (\ref{5.5}) that medium effects on the interacting
hadronic tensor are generated by the energy conserving $\delta$ function, in
that the intrinsic energy available to the pair is fixed by its $CM$
four-momentum, and, therefore, by the momentum distribution $n_{CM}^{N_1
N_2}$ appearing in Eq. (\ref{5.4}); even if the $CM$ motion is neglected
($n_{CM}^{N_1 N_2} = \delta(\vec{k}_{CM})$) \cite{FS}, medium effects still
would remain through the quantity $k_{CM}^0$. We have calculated the
inclusive cross section for the deuteron using the Read Soft Core $NN$
potential \cite{RSC}, taking into account the rescattering in S, P and  D
partial waves; then, using the same two-nucleon amplitudes $\langle
\beta_{12} | j_{\mu}^{N_1} + j_{\mu}^{N_2} | f_{12} \rangle$, we have
computed the cross section $\sigma_1^{(A)}$ for complex nuclei. The results
are shown by the dashed lines in Fig. 14, where the inclusive cross
section for $^2H$, $^3He$, $^4He$ and $^{56}Fe$ are plotted versus the
energy transfer $\nu$ at $Q^2 \simeq 2 ~ (GeV/c)^2$. In Fig. 15 our
results expressed in terms of the nuclear scaling function $F(y,q)$ (see Ref.
\cite{CPS91}) are plotted against the squared three-momentum transfer $q^2$
for fixed values of the scaling variable $y$ (we remind the reader that $y=0$
($<0$) corresponds to $x=1$ ($>1$)). It can be seen that at $1.3 < x < 2$ the
process of two-nucleon rescattering brings theoretical predictions in good
agreement with the experimental data taken from Refs. \cite{SLAC_DEUT} and
\cite{DAY}. The most striking aspect of our results is that the same
mechanism which explains the deuteron data, does the same in a complex
nucleus, provided the A-dependence due to $n_{CM}^{N_1 N_2}$ and $k_{CM}^0$
(clearly exhibited in Fig. 15) is properly considered.

\indent However, it can also be seen from Fig. 14 that the two-nucleon
rescattering is not able to describe the experimental data at $x > 2$.  This
fact is not surprising, since at $x > 2$ more than two nucleons should be
involved in the scattering process. In Ref. \cite{CS94} this process has been
mocked up by considering the motion of the outgoing nucleon, knocked-out from
shell model states, in the complex optical potential generated by the ground
state of the (A-1)-nucleon system. However, the treatment of the
single-nucleon rescattering at $x > 1.3$, based only on the use of on-shell
optical potentials, is not justified (cf. also Refs. \cite{FS}, \cite{FSDS93}
and \cite{UDS89}). Indeed, the struck nucleon, having four-momentum squared
$p'^2 \equiv (\nu + M - E)^2 - (\vec{k} + \vec{q})^2$, can be either
on-mass-shell ($p'^2 = M^2$) or off-mass-shell ($p'^2 \neq M^2$) depending on
the values of $k$ and $E$. Ground-state configurations with $k < |y|$ always
give rise to intermediate off-shell (virtual) nucleons, whose rescattering
amplitudes are expected to decrease with virtuality, for off-shell nucleons
have to interact within short times. Taking into account off-shell effects as
in Ref. \cite{CS94}, our results, including both the single-nucleon and the
two-nucleon rescattering, are presented in Fig. 15. The agreement with the
experimental data is good and holds in the whole low-energy side of the QE
peak. To sum up, both initial state correlations and final state interaction
resemble the ones occurring in the deuteron, apart from the $CM$ motion and the
binding of the pair in a complex nucleus; we have shown that the effects of
$NN$ correlations on the inclusive cross section at $x > 1$ can be described
by applying the factorization hypothesis to the initial as well as to the
final states.

\vspace{1cm}

{\large{\bf 6. $NN$ correlations and exclusive quasi-elastic electron
 scattering}}

\vspace{0.5cm}

\indent The most direct way to check our $NN$ correlation description of the
Spectral Function would be by exclusive experiments, in which either both
$k$ and $E$ are measured (e.g., a $(e,e'N)$ process) or the two nucleons of
the correlated pair are knocked out and detected in coincidence (e.g., a
$(e,e'2N)$ process). We briefly discuss the first process, which, as already
pointed out, has provided a direct evidence of the two-nucleon correlation
model \cite{MAR88,MAG94}. As a matter of fact, within the $IA$ the exclusive
cross section for the process $A(e,e'p)X$ is directly proportional to the
proton Spectral Function through
 \be
    {d^6 \sigma \over dE_{e'} d\Omega_{e'} d\Omega_p dE} = K ~ Z ~
    \sigma_{ep} ~ [P_0(k, E) + P_1(k, E)]
    \label{6.1}
 \ee
where $K = p (M + T_p) {dT_p \over dE_m}$ is a kinematical factor (with
$p=|\vec{p}|$ ($T_p$) being the momentum (kinetic energy) of the detected
proton) and $Z$ is the number of protons. Within the $IA$ one has $k = k_m$
and $E = E_m$, where $k_m$ and $E_m$ are the experimentally measurable
missing momentum ($k_m \equiv |\vec{p} - \vec{q}|$) and missing energy
($E_m \equiv \nu - T_p - T_{A-1}$), respectively. The cross section for fixed
value of $k_m$ will therefore exhibit the shell-model structure given by the
"one-hole" Spectral Function $P_0$  (see Eq. (\ref{2.2.8})), appearing as
peaks corresponding to the single-particle states located at values $E_m =
|\varepsilon_{\alpha}|$ not affected by $k_m$. For large values of $k_m$ ($>
1.5 ~ fm^{-1}$) one is expected to observe the structure generated by the
correlated part $P_1$, i.e. a peak located at $E_m \sim (A-2) / (A-1) ~ k_m^2
/ 2M$, whose position changes with the value of $k_m$. Such a picture has
indeed experimentally observed in the few-nucleon systems, where only one
peak is generated by $P_0$ at $E_m = E_{min}$ (see Eq. (\ref{2.2.12})). The
cross sections for the processes $^3He(e,e'p)X$ and $^4He(e,e'p)X$, measured
in Refs. \cite{MAR88} and \cite{MAG94}at various values of the proton
detection angle $\theta_p$, are presented in Fig. 16, where they are compared
with our predictions within the $IA$ (Eq. (\ref{6.1})) and with the results
of a calculation \cite{LAGET_3,LAGET_4} which also include contributions from
$FSI$ and Meson Exchange Currents ($MEC$). The structure predicted by Eq.
(\ref{6.1}) is clearly seen, i.e. the two-body disintegration peak located at
$E_m \simeq 5.5 ~ MeV$ in $^3He$ ($X =$ $^2H$) and $E_{min} \simeq 19.8 ~
MeV$ in $^4He$ ($X =$ $^3He$)as well as the correlation peak (corresponding
to $X = np$ in $^3He$ and $X = n ^2H, npn$ in $^4He$) located at $E_m \sim
k^2 / 4M$ in $^3He$ and $E_m \sim k^2 / 3M$ in $^4He$. It can be seen that
the presence of $FSI$ and $MEC$, which are clearly relevant at $\theta_p =
90^o$ in $^3He$ and $\theta_p = 134^o$ in $^4He$, only affect the magnitude
of the correlation peak, without sharply modifying its width and location.
Thus, the $A(e,e'N)X$ reaction off few-nucleon system is a direct
confirmation of the validity of our model Spectral Function as well as of the
original $2NC$ model of Ref. \cite{FS} as far as the peak location is
concerned.

\vspace{1cm}

{\large{\bf 7. Summary and conclusions}}

\vspace{0.5cm}

\indent We have demonstrated that at high values of $k$ and $E$ the nucleon
Spectral Function can be expressed as a convolution integral involving the
momentum distributions of the relative and center-of-mass motion of a
correlated nucleon-nucleon pair. The basic step leading to such a
convolution formula is the assumption of the factorization (\ref{3.2.6}) of
the nuclear wave function, which we claim to be the basic configuration
which produces the high momentum and high removal energy part of the
Spectral Function. We have verified {\em a posteriori} the correctness of the
factorization assumption by comparing our model Spectral Function with the
{\em exact} ones in case of the three-nucleon system and infinite nuclear
matter; the main outcome of such a comparison is a very good agreement in the
range of energy and momentum pertaining to the region of two-nucleon
correlations, which is the one investigated in the present paper. As
discussed in Section 4, two-nucleon correlations cover a range of values of
nucleon momentum $k$ and removal energy $E$ characterised by
 \be
    & k & > 2 ~ \mbox{fm$^{-1}$} \nonumber \\
    E_L < & E & < E_{thr}^{(2)} ~ + ~ {A - 2 \over A - 1} ~ {k^2 \over 2M} ~
    [1 - 2 \gamma] ~ + ~ \sqrt{{<k_{CM}^2> ~ 8 ln2 \over 3}} ~ [1 - \gamma] ~
    {k \over M}
    \label{7.1}
 \ee
with $\gamma = {A - 1 \over 2(A - 2)} ~ {<k_{CM}^2> \over <k_{rel}^2>}$ and
$E_L \sim 30 \div 50 ~ MeV$, which leads, e.g. for $k = 2.0, ~ 2.5, ~ 3.0 ~
fm^{-1}$, to $30 ~ MeV < E < 110, ~ 150, ~ 190 ~ MeV$ for $^3He$, and $50 ~
MeV < E < 230, ~ 300, ~ 380 ~ MeV$ for nuclear matter, respectively. Outside
this range, the effects from three-nucleon correlations (i.e., the sharing of
a nucleon high-momentum component by two other nucleons) as well as from
Final State Interaction between the recoiling nucleon and the ($A-2$) system,
is expected to modify the picture predicted by two-nucleon correlations only.
As a matter of fact, the model and exact Spectral Functions of $^3He$
substantially differ in the region of values of $E$ well outside the range
given by Eq. (\ref{7.1}). On the other hand side, in case of infinite nuclear
matter, for which realistic many-body calculations include only ground-state
$2p - 2h$ correlations, so that the three-nucleon continuum states are
absent, the model and the exact Spectral Functions are in very good agreement
in the full range of existing calculations at $k > k_F$. Thus, we are very
confident that the convolution formula realistically describes the high
momentum and high removal energy parts of the nucleon Spectral Function in
complex nuclei generated by two-nucleon correlations.

\indent Using the convolution formula for the Spectral Function we have
calculated the cross section for inclusive and exclusive quasi-elastic
electron scattering off nuclei. The latter process can provide a direct check
of the $k - E$ correlation dependence of the Spectral Function, and available
data do indeed confirm the correctness of the convolution model. As for
inclusive scattering, we have shown that the cross section at $x < 2$ can be
interpreted as due to the coupling of the virtual photon to a correlated
nucleon-nucleon pair and the experimental data can quantitatively be
explained provided the Final State Interaction is taken into account. As a
final remark, we would like to point out that a direct check of the
factorization assumption would provide a conclusive and stringent test of the
validity of the convolution model.

\vspace{1cm}

\indent {\large{\bf Acknowledgements}}

\indent The authors gratefully acknowledge A. Baldo, M. Borromeo, W. Dickhoff,
L.L. Frankfurt, E. Pace, A. Polls, G. Salm\`e, S. Scopetta  and M.I. Strikman
for many fruitful discussions and enlightening remarks during the last years.
Thanks are also due to O. Benhar for supplying us with the numerical output
of his calculations of the nucleon Spectral Function in infinite nuclear
matter.

\newpage

\vspace{1cm}

\indent {\bf Appendix.} Parametrization  of the nucleon momentum distribution
obtained from realistic many-body calculations

\vspace{0.5cm}

\indent In this Appendix a simple parametrization of the   nucleon momentum
distributions $n_0(k)$ (Eq.(\ref{2.3.5})) and $n_1(k)$ (Eq. \ref{2.3.6})),
shown in Fig. 1, is presented. For $n_0(k)$ the following functional forms
have been adopted
 \be
    n_0(k) & = & \sum_{i=1}^{m_0} ~ A_i^{(0)} ~ {e^{-B_i^{(0)} ~ k^2} \over
    (1 + C_i^{(0)} ~ k^2)^2} ~~~~~~~~~~~~~~~~~~~~~~~~~~~~~~~~
    \mbox{for $A = 2, 3, 4$}
    \nonumber \\
    & = & A^{(0)} ~ e^{-B^{(0)} ~ k^2} \cdot \nonumber \\
    & ~ & [1 + C^{(0)} ~ k^2 + D^{(0)} ~ k^4 + E^{(0)} ~ k^6 + F^{(0)}~ k^8]
    ~~~~~~ \mbox{for $4 < A \leq \infty$}
    \label{A.1}
 \ee
The values of the parameters appearing in Eq. (\ref{A.1}) are listed in
Tables A.1 and A.2. For $n_1(k)$, we have adopted a simple two-gaussian
behaviour for all nuclei considered in this paper (but $^3He$), viz.
 \be
    n_1(k) = A^{(1)} ~ e^{-B^{(1)} ~ k^2} + C^{(1)} ~ e^{-D^{(1)} ~ k^2}
    \label{A.2}
 \ee
The values of the parameters appearing in Eq. (\ref{A.2}) are listed in Table
A.3. For $^3He$ the following parametrization nicely reproduces the results
of Ref. \cite{HE3}:
 \be
    n_1(k) = 7.40 ~ {e^{-1.23 ~ k^2} \over (1 + 3.21 ~ k^2)^2} + 0.0139 ~
    e^{-0.234 ~ k^2}
    \label{A.3}
 \ee
with $k$ in $fm^{-1}$ and $n_1(k)$ in $fm^3$. The normalization of $n_0(k)$
and $n_1(k)$ are as in Eq. (\ref{2.3.8}). Finally, it should be pointed out
that the quality of the parametrizations adopted is satisfactory for all
nuclei considered.

\newpage

\begin{center}

{\bf Table Captions}

\end{center}

\vspace{0.5cm}

\indent Table 1. The values of the parameters $C^A$ (Eq.(\ref{4.1.1})) and
$\alpha_{CM}$ (Eq. (\ref{4.1.2})) for various nuclei. The value of $C^A$ is
estimated from the height of the plateaux exhibited by the ratio of the
nucleon momentum distribution $n(k)$ of a nucleus to the one of the deuteron
at $k > 2 ~ fm^{-1}$ (see Fig. 2(b)). The value of $\alpha_{CM}$ is calculated
using Eq. (\ref{4.1.3}). In case of $^3He$ and $^4He$ the value of
$<T>^{(SM)}$ is estimated adopting a simple harmonic oscillator ($HO$) wave
function with the value of the $HO$ length chosen in order to reproduce the
experimental value of the charge radius of the nucleus. In case of $A =
\infty$ the Fermi-gas prediction (i.e., $<T>^{(FG)} = 3/5 k_F^2 / 2M$) is
considered with $k_F = 1.33 ~ fm^{-1}$.

\vspace{0.5cm}

\indent Table 2. Comparison of the mean removal energy in various
nuclei extracted from the energy-weighted sum rule (\ref{4.4.3})
\cite{KOL74} ($<E>$) with the corresponding value calculated within the
extended $2NC$ model of the nucleon Spectral Function ($<E>^{th}$). The
experimental values of the total binding energy per particle
($\varepsilon_A$) and the mean values of the nucleon kinetic energy ($<T>$),
obtained within many-body approaches, are also reported.

\vspace{0.5cm}

\indent Table A.1. Values of the parameters $A_i^{(0)}$, $B_i^{(0)}$ and
$C_i^{(0)}$ appearing in the parametrization (\ref{A.1}) of the nucleon
momentum distribution $n_0(k)$ corresponding to the ground-to-ground
transition (Eq. (\ref{2.3.7})) for few-body systems.

\vspace{0.5cm}

\indent Table A.2. Values of the parameters $A^{(0)}$, $B^{(0)}$, $~C^{(0)}$,
$D^{(0)}$, $E^{(0)}$ and $F^{(0)}$ appearing in the parametrization
(\ref{A.1}) of the "one-hole part" of the nucleon momentum distribution
$n_0(k)$ (Eq. (\ref{2.3.7bis})) for complex nuclei and nuclear matter.

\vspace{0.5cm}

\indent Table A.3. Values of the parameters $A^{(0)}$, $B^{(0)}$, $~C^{(0)}$
and $D^{(0)}$ appearing in the parametrization (\ref{A.2}) of the
"correlated part" of the nucleon momentum distribution $n_1(k)$ (Eq.
(\ref{2.3.6})) for various nuclei.

\newpage

\vspace{0.5cm}

\begin{center}

{\bf Figure Captions}

\end{center}

\vspace{0.5cm}

\indent Fig. 1. The many-body nucleon momentum distribution $n(k)$
(Eq. \ref{2.3.1})) corresponding to the parametrization described in the
Appendix (solid lines). Considering the representation (\ref{2.3.4}),
the momentum distribution $n_0(k)$ (Eq. (\ref{2.3.7}) for $A = 3, 4$ and
Eq. (\ref{2.3.7bis}) for complex nuclei) is given  by the dotted lines.
The deuteron momentum distribution has been calculated using the Paris
potential \cite{PARIS} and the many-body results for the momentum
distributions $n_0(k)$ and $n(k)$ have been taken from Refs.
\cite{HE3} ($^3He$), \cite{HE4_3} ($^4He$), \cite{C12} ($^{12}C$ and
$^{40}Ca$), \cite{O16} ($^{16}O$), \cite{FE56} ($^{56}Fe$), \cite{PB208}
($^{208}Pb$) and \cite{NM} (infinite nuclear matter). The normalization of
$n(k)$ is $\int_0^{\infty} dk ~ k^2 ~ n(k) = 1$. The theoretical calculations
are compared with the experimental values of $n_0(k)$ and $n(k)$ extracted
from the experimental data on inclusive $A(e,e')X$ and exclusive $A(e,e'p)X$
reactions. The open squares represent the results obtained within the
$y$-scaling analysis of inclusive data for $^2H$, $^3He$, $^4He$,$^{12}C$,
$^{56}Fe$ and nuclear matter performed in Ref. \cite{CPS91},  and the full
triangles are the results for $n(k)$ in $^4He$ extracted from the exclusive
reaction $^4He(e,e'p)X$ \cite{MAG94}. The open triangles represent the values
of $n_0(k)$ obtained from the exclusive experiments off $^2H$ \cite{DEUT_p},
$^3He$ \cite{MAR88}, $^4He$ \cite{HE4_p} and $^{12}C$ \cite{C12_p}.

\vspace{0.5cm}

\indent Fig. 2. The nucleon momentum distributions of Fig. 1 shown
all together (a) and their  ratio to the
deuteron momentum distribution $n^{(D)}(k)$ (b). The solid, dashed, dotted,
dot - dashed, long dashed, dot - long dashed lines correspond to  $^2H$,
$^3He$, $^4He$, $^{16}O$, $^{56}Fe$ and nuclear matter, respectively.

\vspace{0.5cm}

\indent Fig. 3. The saturation of the momentum sum rule in $^3He$ (a)
and infinite nuclear matter (b). The dotted and solid lines correspond to the
momentum distribution $n_0(k)$ and to the total momentum distribution $n(k)$,
respectively. In case of $^3He$ the dot-dashed, dashed and long dashed
lines correspond to Eq. (\ref{2.5.1}) calculated in Ref. \cite{HE3} at $E_f =
17.75, 55.5, 305.5 ~ MeV$, whereas for nuclear matter the dot - dashed and
dashed lines correspond \cite{BFF} to $E_f = 100$ and $300 ~ MeV$,
respectively.

\vspace{0.5cm}

\indent Fig. 4. Momentum and removal energy dependences of the nucleon
Spectral Function $P(k,E)$ (times $k^2$) calculated in Ref. \cite{CPS80} for
$^3He$ (a) and in Ref. \cite{BFF} for infinite nuclear matter (b).

\vspace{0.5cm}

\indent Fig. 5. Comparison of the values of the mean removal energy
$<E(k)>_1$ (open squares), defined by Eq. (\ref{3.1.5}), and the peak position
$E_1^{peak}$ (full squares) for $^3He$ and nuclear matter, calculated with
the Spectral Functions of Refs. \cite{CPS80} and \cite{BFF}, respectively,
with the predictions of the $2NC$ model (see Eq. (\ref{3.1.3})). The dotted
lines are the results obtained using Eqs. (\ref{3.1.6}) and (\ref{3.1.4}),
which yield explicitely $<E(k)>_1 = E_1^{peak} = E_{thr}^{(2)} + {k^2 \over 4
M}$ in case of $^3He$ and $<E(k)>_1 = E_1^{peak}  = E_{thr}^{(2)} + {k^2
\over 2 M}$ for nuclear matter.

\vspace{0.5cm}

\indent Fig. 6. The high-momentum part of the relative ($rel$) momentum
distribution and the low-momentum part of Center-of Mass ($CM$) momentum
distribution (see Eq. (\ref{3.2.3bis}) for the definition of the relative
and $CM$ momenta of a correlated $NN$ pair). In case of $^3He$ the open (full)
dots correspond to the results of  Refs. \cite{CPS80} and \cite{FAD}(b),
for a $pp$ ($nn$) pair, respectively (the interaction is  the $RSC$ one
 \cite{RSC}); the full and dashed
lines for $n_{rel}$ represent the rescaled deuteron momentum distributions
(Eq. (\ref{4.1.1}) with $C^A = 1.8$ for the $pp$ pair and $C^A = 2.5$ for the
$pn$ pair), whereas the full line for $n_{CM}$ is the Gaussian parametrization
given by Eq. (\ref{4.1.2}). In case of $^4He$ the triangles represent the
results obtained in Ref. \cite{HE4_1} using the $RSC$ interaction, whereas the
full curves are the rescaled deuteron momentum distribution for $n_{rel}$ (Eq.
(\ref{4.1.1}) with $C^A = 3.2$) and the Gaussian parametrization of Eq.
(\ref{4.1.2}) for $n_{CM}$. The values adopted for the parameter $\alpha_{CM}$
are reported  in Table 1.

\vspace{0.5cm}

\indent Fig. 7. Comparison of the values of the mean removal energy $<E(k)>_1$
(open squares), defined by Eq. (\ref{3.1.5}), and the peak position
$E_1^{peak}$ (full squares) for $^3He$ and nuclear matter, calculated with
the Spectral Functions of Refs. \cite{CPS80} and \cite{BFF}, respectively,
with the predictions of the extended $2NC$ model (see Eq. (\ref{3.2.12})). The
dashed and solid lines correspond, respectively, to the values of $<E(k)>_1$
and $E_1^{peak}$ obtained using in Eq. (\ref{3.2.14}) the effective relative
and $CM$ momentum distributions given by Eqs. (\ref{4.1.1}) and (\ref{4.1.2}),
respectively.

\vspace{0.5cm}

\indent Fig. 8. Comparison of the values of the Full Width @ Half Maximum
(FWHM) calculated using the Spectral Functions of Refs. \cite{CPS80} for
$^3He$ (open dots) and \cite{BFF} for nuclear matter (open squares), with the
predictions of the extended $2NC$ model (see Eq. (\ref{3.2.12}). The dashed
line corresponds to the predictions of Eq. (\ref{4.2.3}), whereas the solid
line is the results obtained using our model Spectral Function, calculated
using in Eq. (\ref{3.2.14}) the effective relative and $CM$ momentum
distributions given by Eqs. (\ref{4.1.1}) and (\ref{4.1.2}), respectively.
The dotted line represents the prediction of the "naive" $2NC$ model (see
Eq. (\ref{3.1.3}), in which the $CM$ of the correlated pair is assumed to be
at rest.

\vspace{0.5cm}

\indent Fig. 9. The nucleon Spectral Function of $^3He$ \cite{CPS80} and
nuclear matter \cite{BFF} versus the removal energy $E$ for various values of
the nucleon momentum. For $^3He$ (a) the squares, full dots and open dots
correspond to $k = 2.2, 2.8$ and $3.5 ~ fm^{-1}$, respectively. For
nuclear matter (b) the open squares, full dots, open dots and full
squares correspond to $k = 1.5, 2.2, 3.0$ and $3.5 ~ fm^{-1}$, respectively.
The solid lines are the predictions of our extended $2NC$ model obtained
using in Eq. (\ref{3.2.14}) the effective relative and $CM$ momentum
distributions given by Eqs. (\ref{4.1.1}) and (\ref{4.1.2}). The value of
the constant ${\cal{N}}$ appearing in Eq. (\ref{3.2.12}) is fixed by Eq.
(\ref{3.2.13}), in which the correlated part $n_1(k)$ of the nucleon
momentum distribution calculated in Refs. \cite{CPS80} and \cite{BFF} has
been used.

\vspace{0.5cm}

\indent Fig. 10. Momentum and removal energy dependences of the nucleon
Spectral Function $P(k,E)$ (times $k^2$) for $A = 3$ (left side) and $A =
\infty$ (right side). The predictions of our extended $2NC$ model, obtained
using in Eq. (\ref{3.2.14}) the effective relative and $CM$ momentum
distributions given by Eqs. (\ref{4.1.1}) and (\ref{4.1.2}), are shown in
the lower part of the figure, whereas the results of the many-body
calculations of Refs. \cite{CPS80} and \cite{BFF} are shown in the upper
one.

\vspace{0.5cm}

\indent Fig. 11. The nucleon Spectral Function of $^4He$ \cite{MS91} versus
the removal energy $E$ for two values of the nucleon momentum: $k = 3 ~
fm^{-1}$ (a) and $k = 4 ~ fm^{-1}$ (b). The open squares are the results of
the many-body calculations of Ref. \cite{MS91}, whereas the solid lines are
the predictions of Eq. (\ref{3.2.14}) using the effective relative and $CM$
momentum distributions given by Eqs. (\ref{4.1.1}) and (\ref{4.1.2}).
The dot-dashed, dotted and dashed lines are the results of Eq. 61
obatained using for $w(|\vec{t}|)$ the hard part of the deuteron momentum
distribution and adopting the values $1.0, 1.5$ and $2.0 ~ fm^{-1}$,
respectively, for the upper limit of integration over $|\vec{t}| = \sqrt{M
E_2^*}$, with $E_2^*$ being the (positive) excitation energy of the
residual two-nucleon system.

\vspace{0.5cm}

\indent Fig. 12. Momentum and removal energy dependences of the nucleon
Spectral Function $P(k,E)$ (times $k^2$) predicted by our extended $2NC$
model for $^{16}O$ using in Eq. (\ref{3.2.14}) the effective relative and $CM$
momentum distributions given by Eqs. (\ref{4.1.1}) and (\ref{4.1.2}).

\vspace{0.5cm}

\indent Fig. 13. The saturation of the energy ($S_f(E)$, (\ref{4.4.1})) and
momentum ($n_f(k)$, Eq. (\ref{4.4.2})) sum rules calculated for $^{16}O$ and
$^{208}Pb$ within our extended $2NC$ model. The dotted and solid lines
represent the values of $S_f(E)$ and $n_f(k)$ obtained by considering in Eqs.
(\ref{4.4.1}-\ref{4.4.2}) the limits $k_f$ ($E_f$) $\rightarrow \infty$ and
using the one-hole $P_0(k,E)$ and the total $P_0(k,E) + P_1(k,E)$ Spectral
Functions, respectively. As for $S_f(E)$, the long-dashed, dashed and
dot-dashed lines correspond to Eq. (\ref{4.4.1}) calculated at $k_f =
1.5, 2.0$ and $3.0 ~ fm^{-1}$, whereas in case of $n_f(k)$ they
represent the results of Eq. (\ref{4.4.2}) obtained at $E_f = 50, 100$
and $300 ~ MeV$. For the calculation of $P_0(k,E)$ Eq. (\ref{2.2.8}) has
been used, with the value of the shell-model parameters taken from Ref.
\cite{C12_p}.

\vspace{0.5cm}

\indent Fig. 14. Inclusive cross section $\sigma(Q^2, \nu)$ (\ref{5.1}) for
the process $A(e,e')X$ versus the energy transfer $\nu$ at $Q^2 \simeq 2 ~
(GeV/c)^2$. The values of the Bjorken variable $x$ are reported in the
upper axis. The experimental data are taken from Refs. \cite{SLAC_DEUT} and
\cite{DAY}. Calculations have been performed using the free nucleon form
factors of Ref. \cite{GAL71}, the $cc1$ prescription of Ref. \cite{DEF83} for
$\sigma_{eN}$ and the $RSC$ potential \cite{RSC} for the $NN$ interaction.
Dotted lines: Impulse Approximation (Eqs. (\ref{5.2}) and (\ref{5.3}))
obtained using our extended $2NC$ model for the nucleon Spectral Function.
Dashed lines: $IA$ + two-nucleon rescattering \cite{CS94}. Dot-dashed lines:
contribution from nucleon inelastic channels estimated as in Ref. \cite{CS94}.

\vspace{0.5cm}

\indent Fig. 15. Nuclear scaling function $F(y,q)$ for $^2H$, $^4He$
and $^{56}Fe$ versus the squared three-momentum transfer $q^2$ for
fixed values of the scaling variable $y$ \cite{CPS91}. The values of the
Bjorken variable $x$ are shown in the upper axis. Dotted lines: Impulse
Approximation obtained using our extended $2NC$ model of the nucleon
Spectral Function; dashed lines: correlated $NN$ pair contribution;
dot-dashed lines: {\em one-hole} contribution; solid line: $IA$ + full final
state interaction \cite{CS94}.

\vspace{0.5cm}

\indent Fig. 16. Exclusive cross sections for the processes
$^3He(e,e'p)X$ and $^4He(e,e'p)X$ versus the missing energy $E_m$ for
various values of the detection angle $\theta_p$ of the proton. The solid
histograms show the data of Ref. \cite{MAR88} for $^3He$ and Ref. \cite{MAG94}
for $^4He$ after radiative corrections. Dotted lines: Impulse Approximation
(Eq. \ref{6.1}) calculated using our extended $2NC$ model for the nucleon
Spectral Function and adopting the free nucleon form factors of Ref.
\cite{GAL71} and the $cc1$ prescription of Ref. \cite{DEF83} for
$\sigma_{ep}$. Dashed lines: distorted wave impulse approximation + meson
exchange currents calculated in Ref. \cite{LAGET_3} for $^3He$ and in Ref.
\cite{LAGET_4} for $^4He$. The arrows locate the values of $E_m$
corresponding to the maxima of the cross section expected within the $IA$;
the corresponding values of the missing momentum $k_m$ are also
reported. Note that within the $IA$ one has $k = k_m$ and $E = E_m$.

\newpage

\begin{center}

\vspace{3cm}

{\bf TABLE 1}

\vspace{1cm}

\begin{tabular}{||c ||c ||c |c ||} \hline
 $Nucleus$ & $ ~ C^A ~ $ & $<T>^{(SM)}$ & $\alpha_{CM}$     \\

           &             &   $(MeV)$    &   $(fm^2)$        \\ \hline

    $^3He$ &     1.9     &     8.5      &     3.7           \\ \hline

    $^4He$ &     3.8     &     9.8      &     2.4           \\ \hline

  $^{12}C$ &     4.0     &    16.9      &     1.0           \\ \hline

  $^{16}O$ &     4.2     &    14.0      &     1.2           \\ \hline

 $^{40}Ca$ &     4.4     &    16.7      &     0.98          \\ \hline

 $^{56}Fe$ &     4.5     &    14.3      &     1.1           \\ \hline

$^{208}Pb$ &     4.8     &    18.4      &     0.85          \\ \hline

$A=\infty$ &     4.9     &    22.0     &      0.71          \\ \hline
\end{tabular}

\end{center}

\newpage

\begin{center}

\vspace{3cm}

{\bf TABLE 2}

\vspace{1cm}

\begin{tabular}{||c ||c |c |c ||c ||} \hline
 $Nucleus$ & $\varepsilon_A$ &  $<T>$  &  $<E>$  & $<E>^{th}$   \\

           &     $(MeV)$     & $(MeV)$ & $(MeV)$ &   $(MeV)$    \\ \hline

  $^{12}C$ &      -7.7       &  32.4   &   44.9  &    44.8      \\ \hline

  $^{16}O$ &      -8.0       &  30.9   &   44.8  &    46.5      \\ \hline

 $^{40}Ca$ &      -8.5       &  33.8   &   50.1  &    49.6      \\ \hline

 $^{56}Fe$ &      -8.8       &  32.7   &   49.7  &    48.9      \\ \hline

$^{208}Pb$ &      -7.9       &  38.2   &   53.7  &    51.8      \\ \hline
\end{tabular}

\end{center}

\newpage

\begin{center}

\vspace{3cm}

{\bf TABLE A.1}

\vspace{1cm}

\begin{tabular}{||c ||c ||c ||c ||} \hline
     $Nucleus$      &   $^2H$   &   $^3He$   &   $^4He$    \\

                    &           & $(proton)$ &             \\ \hline

 $A_1^{(0)} (fm^3)$ & 157.4     &    31.7    &    4.33     \\ \hline

 $B_1^{(0)} (fm^2)$ &   1.24    &    1.32    &    1.54     \\ \hline

 $C_1^{(0)} (fm^2)$ &  18.3     &    5.98    &     0.419   \\ \hline

 $A_2^{(0)} (fm^3)$ &   0.234   &    0.00266 &     5.49    \\ \hline

 $B_2^{(0)} (fm^2)$ &   1.27    &    0.365   &     4.90    \\ \hline

 $C_2^{(0)} (fm^2)$ &    --     &     --     &      --     \\ \hline

 $A_3^{(0)} (fm^3)$ &   0.00623 &     --     &      --     \\ \hline

 $B_3^{(0)} (fm^2)$ &   0.220   &     --     &      --     \\ \hline

 $C_3^{(0)} (fm^2)$ &     --    &     --     &      --      \\ \hline
\end{tabular}

\end{center}

\newpage

\begin{center}

\vspace{3cm}

{\bf TABLE A.2}

\vspace{1cm}

\begin{tabular}{||c ||c |c |c |c |c |c ||} \hline
 $Nucleus$ & $A^{(0)}$ & $B^{(0)}$ & $C^{(0)}$ & $D^{(0)}$ & $E^{(0)}$ &
 $F^{(0)}$                                                   \\

           & $(fm^3)$  & $(fm^2)$  & $(fm^2)$  & $(fm^4)$  & $(fm^6)$  &
 $(fm^8)$                                                    \\ \hline

  $^{12}C$ &   2.61    &    2.66   &    3.54   &--&--&--     \\ \hline

  $^{16}O$ &   2.74    &    3.33   &    6.66   &--&--&--     \\ \hline

 $^{40}Ca$ &   3.24    &    3.72   & &    11.1 &--&--        \\ \hline

 $^{56}Fe$ &   3.57    &    4.97   & &    19.8   & 15.0 &-   \\ \hline

$^{208}Pb$ &   1.80    &    4.77   & &    25.5   & & 40.3    \\ \hline

$A=\infty$ & & & & & & $ $                                   \\

$(k<k_F)$  &   1.08    &     0.118 &--&--&--&--              \\ \hline
\end{tabular}

\end{center}

\newpage

\begin{center}

\vspace{3cm}

{\bf TABLE A.3}

\vspace{1cm}

\begin{tabular}{||c ||c |c |c |c ||} \hline
 $Nucleus$ & $A^{(1)}$ & $B^{(1)}$ & $C^{(1)}$ & $D^{(1)}$   \\

           & $(fm^3)$  & $(fm^2)$  & $(fm^3)$  & $(fm^2)$    \\ \hline

    $^4He$ &   0.665   &    2.15   &    0.0244 &   0.22      \\ \hline

  $^{12}C$ &   0.426   &    1.60   &    0.0237 &   0.22      \\ \hline

  $^{16}O$ &   0.326   &    1.40   &    0.0263 &   0.22      \\ \hline

 $^{40}Ca$ &   0.419   &    1.77   &    0.0282 &   0.22      \\ \hline

 $^{56}Fe$ &   0.230   &    1.20   &    0.0286 &   0.22      \\ \hline

$^{208}Pb$ &   0.275   &    1.01   &    0.0304 &   0.22      \\ \hline

$A=\infty$ & & & & $ $                                       \\

 $(k<k_F)$ &   0.859   &    0.043  &   -0.839  &   0.12      \\ \hline

$A=\infty$ & & & & $ $                                       \\

 $(k>k_F)$ &   0.432   &    0.97   &    0.0313 &   0.22      \\ \hline
\end{tabular}

\end{center}

\end{document}